\documentclass[journal=mamobx,manuscript=article]{achemso}
\usepackage[version=3]{mhchem} 
\usepackage{amsmath,amssymb}
\usepackage{graphicx}
\usepackage[colorlinks=true,linkcolor=blue,citecolor=blue,urlcolor=blue]{hyperref}
\usepackage{bm}
\usepackage{color}
\usepackage{ulem}
\usepackage{flafter}

\SectionNumbersOn

\newcommand{\doi}[1]{\href{http://dx.doi.org/#1}{\nolinkurl{#1}}}

\title[Topological transition]{
Topological transition in multicyclic chains with structural symmetry inducing stress-overshoot phenomena in multicyclic/linear blends under biaxial elongational flow}

\author{Takahiro Murashima}
\email{murasima@cmpt.phys.tohoku.ac.jp}
\phone{+81-22-795-5718}
\fax{+81-22-795-6447}
\affiliation{Department of Physics, Tohoku University, 6-3, Aramaki-aza-Aoba, Aoba-ku, Sendai, 980-8578, Japan}

\author{Katsumi Hagita}
\affiliation{Department of Applied Physics, National Defense Academy, 1-10-20, Hashirimizu, Yokosuka, 239-8686, Japan}

\author{Toshihiro Kawakatsu}
\affiliation{Department of Physics, Tohoku University, 6-3, Aramaki-aza-Aoba, Aoba-ku, Sendai, 980-8578, Japan}


\abbreviations{IR,NMR,UV}

\begin{document}


\begin{center}
\vspace*{20pt}
\textbf{Graphical TOC Entry}
\vspace*{10pt}

\fbox{
\includegraphics[width=8.3cm]{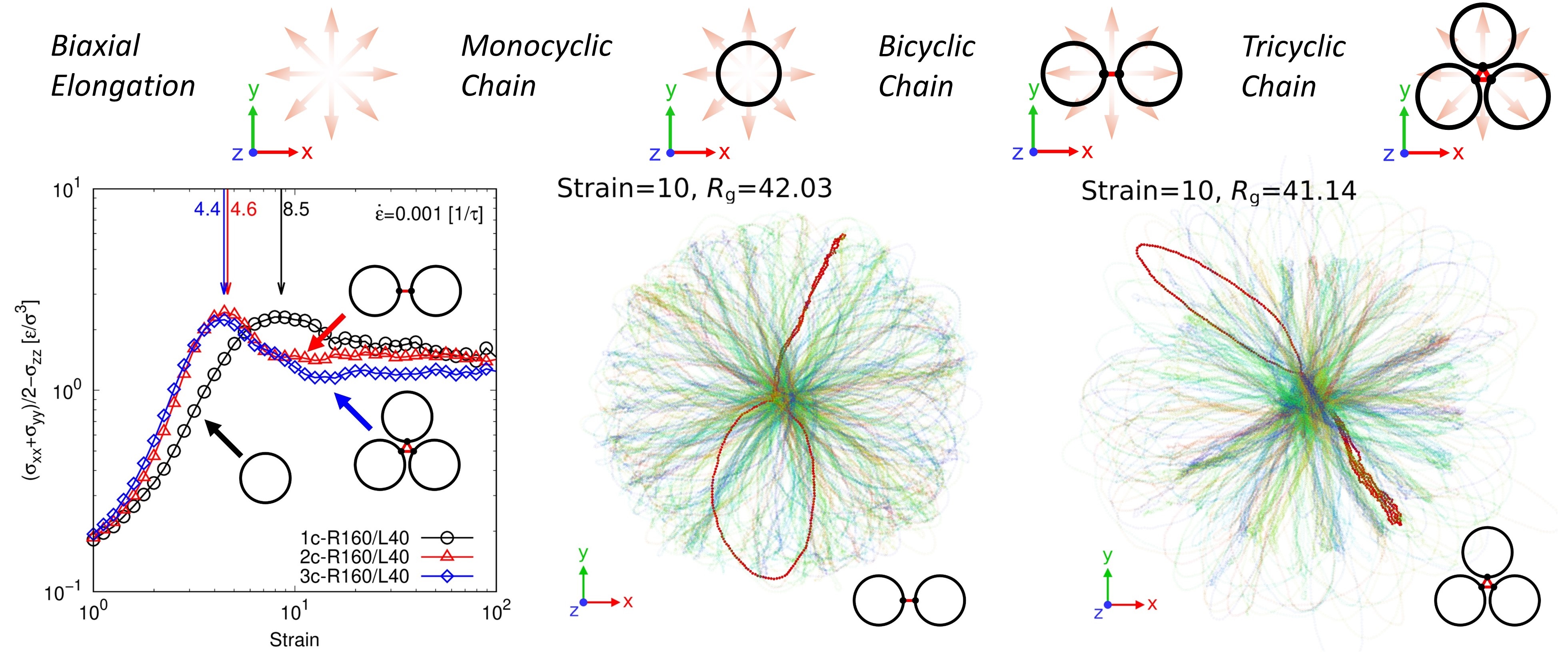}
}

(for Table of Contents use only)
\end{center} 


\clearpage

\begin{abstract}
Blends of multicyclic and linear polymers under biaxial elongational flow were analyzed using coarse-grained molecular dynamics simulations. The multicyclic/linear blends displayed overshoot in the normal stress difference at the start-up of biaxial elongational flow. This overshoot was steeper for multicyclic/linear blends than for our previously reported monocyclic/linear blends [{\it T. Murashima, K. Hagita, and T. Kawakatsu, Macromolecules, \textbf{2021}, 54, 7210}]. Investigation of the origin of the overshoot in the multicyclic/linear blends revealed a different mechanism than that previously observed in our monocyclic/linear blends. Specifically, a “{\it topological transition}” mechanism comprising a morphological change from the open- to closed-ring state was observed in the multicyclic chains, but not in the monocyclic chains. This topological transition drastically changes the stress of the rings. Although the topological transition was also observed in asymmetric-multicyclic/linear blends, no stress overshoot appeared, owing to the asymmetry in the multicyclic chains. Therefore, we hypothesized that the structural symmetry in multicyclic chains is indispensable for overshoot behavior to occur. We determined that the topological transition in multicyclic chains with structural symmetry induces stress-overshoot behavior in multicyclic/linear blends under biaxial elongational flow.
\end{abstract}

\clearpage

\section{Introduction}
Advances in synthetic technology have led to the development of materials with various topological structures that combine multiple rings~\cite{Tezuka2012,JiaMonteiro2012,IsonoEtal2018}.
Such topologies differ from those of conventional linear chains, branched chains, and single rings, resulting in new material functionalities. For example, Hossain et al. observed that the topologies of multicyclic polymers have a marked influence on their glass transition temperature~\cite{HossainEtal2014}.
Yan et al. reported that multicyclic polymers display hierarchical inward relaxation, and that the zero-shear viscosity decreased with an increasing number of constrained segments on the coupling sites~\cite{YanEtal2018}.
Doi et al. investigated a dumbbell-shaped polymer consisting of two small rings and a short linear chain. Their blend of dumbbell-shaped and long linear polymers revealed that the dumbbell chain acts as a pseudo-entanglement point with a longer characteristic time than that of the host linear chain~\cite{DoiEtal2022}.
However, while the understanding of monocyclic rings has progressed rapidly in recent years, the basic understanding of multicyclic polymers, except for these pioneering works, is still in its infancy. This is because many studies have concentrated on monocyclic rings, especially those with much larger molecular weights than the entanglement molecular weight of linear polymers.

Single-ring polymers present different rheological properties than those of linear chains because of the absence of free ends~\cite{McLeish2002}.
Thus, the stress relaxation of ring melts exhibits power-law behavior without an entanglement plateau~\cite{KapnistosEtal2008,HalversonEtal2011}.
Moreover, the ratio of the zero-shear viscosity of a linear polymer melt to that of its ring counterpart increases with the increasing number of entanglements~\cite{PasquiroEtal2013}.
The relaxation dynamics of rings surrounded by obstacles is understood through three typical ring structures~\cite{Klein1986}: (i) obstacle-enclosing, (ii) non-obstacle-enclosing but ramified with double-folded large loops, and (iii) non-obstacle-enclosing and non-ramified (or ramified with small loops).
The constraint release and reptation relaxation mechanisms of the first and third structures are similar to those of linear chains~\cite{DoiEdwards1986}. On the other hand, the relaxation mechanism of the second structure is difficult to understand in analogy with the conventional linear chain dynamics. Instead, similar to an amoeba, such rings invade between the surrounding obstacles while stretching and shrinking many branched double-folded loops, a behavior described by the lattice animal model~\cite{ObukhovEtal1994}. Ge et al. developed the fractal loopy globule model, implementing the self-similar dynamics of entangled rings with self-consistent multi-ring dynamics, and successfully reproduced the simulation and experimental results~\cite{GeEtal2016}.


Rings are penetrated and interlocked by the surrounding rings~\cite{SuburamanianShanbhag2008,LoTurner2013,BernabeiEtal2013,TsalikisEtal2016,SmrekEtal2019}, while in linear chain contaminants they are penetrated by linear chains~\cite{SuburamanianShanbhag2008,TsalikisMavrantzas2014}. Thus, the constraint release mechanism of rings is much slower than theoretically expected. Moreover, since these penetrations lead to slow ring polymer dynamics~\cite{TsalikisMavrantzas2020,MoEtal2022}, the rheological properties of ring/linear blends differ from those of both the pure ring and linear melts. Specifically, a small number of linear chains drastically increases the viscosity of the ring melts~\cite{Roovers1988,HalversonEtal2012}, whereas a low ring volume fraction increases the viscosity of the linear melt up to the overlapping concentration of the rings~\cite{ParisiEtal2020}. In a flow field, rings represent nonlinear behaviors that are different from those of linear chains owing to penetration. Zhou et al. observed transient stretching of the DNA ring in planar elongational flow~\cite{ZhowEtal2019,ZhowEtal2021}, while Borger et al. discovered a threading–unthreading transition in ring/linear blends under uniaxial elongational flow~\cite{BorgerEtal2020}. Parisi et al. investigated the steady-state shear viscosity of pure ring melts and explained their shear thinning behavior using a shear slit model~\cite{ParisiEtal2021a}. They also investigated the shear and elongational viscosities of ring/linear blends~\cite{ParisiEtal2021b}.

Recently, we actively investigated ring/linear blend systems from the viewpoint of penetration control for device application. By defining the number of penetrations using the Gauss linking number~\cite{DoiEdwards1986}, we established the minimum ring size in which the ring has more than one penetration for monocyclic/linear blends through coarse-grained molecular dynamics simulation~\cite{HagitaMurashima2021a}. For multicyclic/linear blends, the number of penetrations is increased by increasing the number of rings~\cite{HagitaMurashima2021b}. Notably, these equilibrated simulation data are publicly available for the development of a broad scientific community~\cite{HagitaEtal2022a}. Following a mapping procedure~\cite{EveraersEtal2020} from the coarse-grained model to a real material, we successfully reproduced the experimental observations~\cite{IwamotoEtal2018}. Moreover, we confirmed that increasing the ring concentration reduces the ring size and number of penetrations owing to the topological repulsion between the rings~\cite{HagitaMurashima2021c}. Cross-linking among the linear chains in a ring/linear blend can create a material in which the rings are trapped in the networks. Interpreting the number of penetrations as the trapping probability of the rings, we confirmed that the trapping probabilities obtained from the experiments and simulations were in good agreement~\cite{HagitaEtal2022b}. Moreover, the trapping probability of multicyclic chains were evaluated, and the tricyclic chains in the network presented different topologies depending on the penetration mode. These topologies were expected to contribute to the physical characteristics of the network. We also studied the role of rings trapped in a network under tensile loading~\cite{HagitaEtal2022c,HagitaEtal2022d}.

To determine the nonlinear rheological properties, elongational and shear flow simulations have been in high demand in recent years. Elongational flow simulation with large strains, however, has remained difficult for many years, except for planar elongational flow, in which the Kraynik--Reinelt (KR) boundary conditions~\cite{KraynikReinelt1992} are available to avoid the collapse of the simulation box at large strains~\cite{BaranyaiCummings1999,MatinEtal2000}. Indeed, the discovery of the generalized Kraynik--Reinelt (gKR) boundary conditions~\cite{Dobson2014,Hunt2015,NicholsonRutledge2016} has seen the progress in the simulation studies of uniaxial/biaxial elongational flows~\cite{MurashimaEtal2018,OConnorEtal2018,OConnorEtal2019}. In particular, when studying ring polymers, O’Connor et al. discovered anomalous thickening behaviors for pure ring melts under uniaxial elongational flow~\cite{OConnorEtal2020}. Further, they systematically investigated a threading–unthreading transition~\cite{BorgerEtal2020} for ring/linear blends with varying ring fractions~\cite{OConnorEtal2022}. Moreover, our group reported stress-overshoot phenomena under biaxial elongational flow for ring/linear blends~\cite{MurashimaEtal2021}.

In our previous study~\cite{MurashimaEtal2021}, we investigated ring/linear blends using the Kremer--Grest (KG)-type bead-spring model~\cite{KremerGrest1990}, where the ring-to-linear chain blend weight ratio was fixed at 1:10. We observed stress-overshoot phenomena under biaxial elongational flow at a ring size, $N_{\rm R}$, of 160 ($N_{\rm R} >$80 --- A ring with $N_{\rm R}=80$ has more than one penetration~\cite{HagitaMurashima2021a}) and linear chain length values, $N_{\rm L}$, of 20, 40, and 80. In contrast, no overshoot was observed at $N_{\rm L}$ values of 10 and 160. The overshoot was apparent at $N_{\rm L}=40$ under biaxial elongational flow, with a strain rate $\dot{\varepsilon}$ of 0.001/$\tau$, where $\tau$ is the unit of time in the Lennard-Jones (LJ) particle system. This strain rate is faster than 1/$\tau_{\rm e}$, where $\tau_{\rm e}$($\approx$ 2000$\tau$~\cite{KremerGrest1990}) is the entanglement time of the KG model. This overshoot behavior originated from excessive ring elongation and slight shrinkage under fast elongational flow. The excessive ring elongation was caused by the linear chains penetrating the ring, whereby the linear chains spread inside the ring, and the ring was pushed outward by the linear chains. When the linear chains penetrating the ring were pulled out, the ring lost the force from the linear chains and subsequently shrank. Excessive ring elongation was suppressed at $N_{\rm L}$ values $>$80, owing to entanglements among the linear chains. For $N_{\rm L}$ values $<$20, the linear chains penetrating the ring detached immediately under the fast elongational flow, and they did not cause the excessive ring elongation. Owing to our limited computational resources, we could not clarify the $N_{\rm R}$-dependence of the stress-overshoot phenomena in our previous work. Furthermore, multicyclic chains with $n$ rings exhibit more penetrations than do monocyclic chains~\cite{HagitaMurashima2021b}. Thus, we predicted that the stress-overshoot phenomena would display the $n$-dependence in multicyclic/linear blends.

Our research group is also interested in the shape of multicyclic chains in the steady state under biaxial elongational flow. As shown in our previous work~\cite{MurashimaEtal2021}, the rings spread in the elongational plane, so that their steady state was the open state. For multicyclic chains, the rings are also expected to spread in the elongational plane, although the shape is thought to depend on the connection of the rings. Both bond-connected and catenated multicyclic chains can be considered~\cite{HagitaMurashima2021b}. For a catenated multicyclic chain (catenane), rings interpenetrate each other and do not have bond connections between them. Because the rings in a catenane can slide and overlap with each other under biaxial elongational flow, the steady-state shape will be close to that of monocyclic chains, i.e., the open state. For a bond-connected multicyclic chain, a junction part connects the rings. The steady state of such a chain under biaxial elongational flow is significant, and the role of the junction should be clarified. Therefore, in the present work, we focused on multicyclic chains with bond connections between the rings.

In this study, we employed coarse-grained molecular dynamics simulations to clarify the stress-overshoot phenomena in multicyclic/linear blends and investigate the $N_{\rm R}$- and $n$-dependences of these phenomena under biaxial elongational flows. In particular, we considered symmetric bicyclic and tricyclic chains with bond connections. Thus, we elucidated a new mechanism that induces stress overshoot in multicyclic/linear blends, which is significantly different from that observed in monocyclic/linear blends. This observed {\it topological transition} mechanism comprises a morphological change in multicyclic chains, which drastically changes the stresses on the rings in multicyclic/linear blends. Herein, we describe the simulation methods and results of our study. Consequently, we reveal that the stress-overshoot phenomena in multicyclic/linear blends under biaxial elongational flow are induced by topological transitions in multicyclic chains with structural symmetry.

\clearpage

\section{Methods}
\begin{figure}[htbp]
    \centering
    \includegraphics[width=7in]{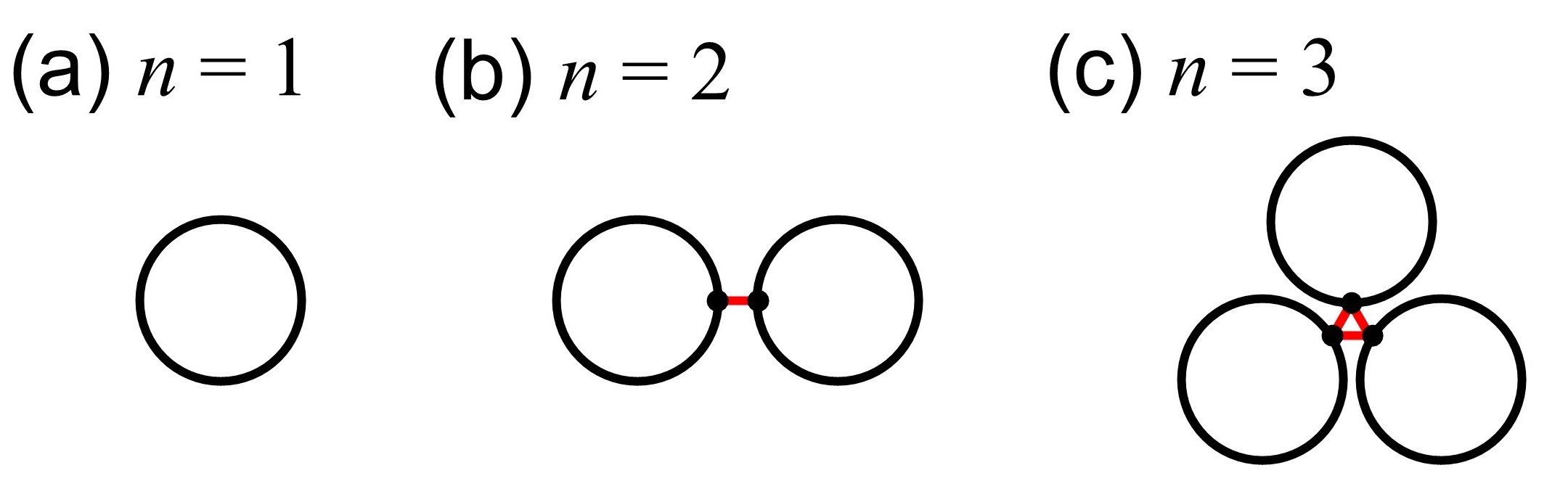}
    \caption{Schematics of (a) monocyclic ($n$ = 1), (b) bicyclic ($n$ = 2), and (c) tricyclic ($n$ = 3) chains. The black circles represent rings, red lines represent extra bonds connecting the rings, and black points represent bond-connected particles. Each black circle consists of $N_{\rm R}$ particles with $N_{\rm R}$ bonds, whereas each red line represents a bond connecting two black points.}
    \label{fig1_schematics}
\end{figure}

We investigated blends of multicyclic and linear chains. The polymer chains considered in the present study are described using the KG model~\cite{KremerGrest1990}: A multicyclic chain consists of $n$ rings, each ring composed of $N_{\rm R}$ particles and $N_{\rm R}$ bonds, while a linear chain consists of $N_{\rm L}$ particles and $N_{\rm L} - 1$ bonds. Thus, a multicyclic chain of $n$ rings comprises $nN_{\rm R}$ particles. For multicyclic chains with $n >$1, an extra bond is required to connect the rings. Following our previous work~\cite{HagitaMurashima2021b}, we considered a bicyclic chain ($n = 2$) to comprise one extra bond connecting two rings and a tricyclic chain ($n = 3$) to comprise three extra bonds connecting three rings, as shown in Figure \ref{fig1_schematics}. For comparison, a monocyclic chain ($n = 1$)~\cite{MurashimaEtal2021} was also considered. When the numbers of rings and linear chains in the system are $M_{\rm R}$ and $M_{\rm L}$, the total number of particles in the system is $N_{\rm R}M_{\rm R} + N_{\rm L}M_{\rm L}$. To fix the number of rings $M_{\rm R}$ in the system at the same $N_{\rm R}$, the number of multicyclic chains was set to $M_{\rm R}/n$. Periodic boundary conditions were applied to the LJ particle system with units of energy $\epsilon$, length $\sigma$, and mass $m$. The time unit in this system was defined as $\tau=\sigma(m/\epsilon)^{1/2}$. To realize uncrossable chains, the repulsive LJ potential with the cutoff length $r_{\rm c} = 2^{1/6}\sigma$ was applied to each pair of particles, and the finite extensible nonlinear elastic (FENE) potential, with spring constant $k = 30\epsilon/\sigma^2$ and maximum bond extent $R_0 = 1.5\sigma$, was applied between the bonded particles. The density $\rho$ was set to 0.85$m/\sigma^3$. The temperature --- $T = 1.0\epsilon/k_{\rm B}$, where $k_{\rm B}$ is the Boltzmann constant --- was controlled by the Langevin thermostat with friction coefficient $\zeta=0.5m/\tau$. The positions and velocities of the particles were time-integrated using the velocity-Verlet algorithm with the time step $\Delta t=0.01\tau$. Hereafter, the unit symbols ($\epsilon$, $\sigma$, $m$, and $\tau$) are not presented for simplicity if there is no confusion. Under a deformation field with deformation rate tensor $\boldsymbol{\kappa}$, the particle positions and velocities were evolved using the SLLOD algorithm~\cite{EvansMorriss2008}. The deformation rate tensor $\boldsymbol{\kappa}$ under biaxial elongational flow with a constant Hencky strain rate $\dot{\varepsilon}$ is expressed as $\boldsymbol{\kappa}$ = diag($\dot{\varepsilon}$, $\dot{\varepsilon}$, $- 2\dot{\varepsilon}$). On applying an elongational deformation to the system for a long period, the simulation box collapses under the conventional method, in which the simulation box is placed parallel to the flow direction. Therefore, to avoid such a collapse, we applied the gKR boundary conditions~\cite{KraynikReinelt1992,Dobson2014,Hunt2015} to the system under elongational flow.

Following our previous work~\cite{MurashimaEtal2021}, the multicyclic-to-linear chain blend weight ratio was set to 1:10 ($(nN_{\rm R})(M_{\rm R}/n)$:$N_{\rm L}M_{\rm L}$ = $N_{\rm R}M_{\rm R}$:$N_{\rm L}M_{\rm L}$), wherein the concentration of the rings is less than the overlap concentration at equilibrium. Because the $N_{\rm R}$-dependence was not considered in our previous work~\cite{MurashimaEtal2021}, we reconsidered the monocyclic/linear blends and compared them with the multicyclic/linear blends. In the present work, we considered 60 cases originating from the combinations of $n$, $(N_{\rm R}, M_{\rm R})$, and $(N_{\rm L}, M_{\rm L})$, where $n$ = \{1, 2, 3\}; $(N_{\rm R} , M_{\rm R})$ = \{(40, 1536), (80, 768), (120, 512), (160, 384)\}; and $(N_{\rm L} , M_{\rm L})$ = \{(10, 61440), (20, 30720), (40, 15360), (80, 7680), (160, 3840)\}. Because such an exhaustive study requires vast computational resources, it is difficult to investigate the various flow conditions. The present work focuses on stress-overshoot phenomena under biaxial elongational flow with $\dot{\varepsilon}$ = 0.001$/\tau$, where monocyclic/linear blends exhibit overshoot behavior for $\dot{\varepsilon} \tau_{\rm e} >$1~\cite{MurashimaEtal2021}. To prepare an equilibrium state for each blend, we used HOOMD-blue~\cite{AndersonEtal2020} with GPU acceleration. Moreover, product runs under biaxial elongational flow were performed using LAMMPS~\cite{Plimpton1995} with the UEFEX package~\cite{MurashimaEtal2018}, which extends the UEF package~\cite{NicholsonRutledge2016} applicable to the Langevin thermostat. Graphical analyses were performed using OVITO~\cite{Stukowski2010}. Sufficiently equilibrated states were produced in our previous studies~\cite{HagitaMurashima2021a,HagitaMurashima2021b,HagitaEtal2022a}, and thus, we used these equilibrated data as the initial conditions.

\clearpage

\section{Results and Discussion}

\subsection{Stress-overshoot behavior under biaxial elongational flow}
This section summarizes the time-evolution behavior of the first normal stress difference $(\sigma_{xx}+\sigma_{yy})/2 - \sigma_{zz}$, which is a typical characteristic variable under biaxial elongational flow. Here, $x$ and $y$ are the elongational directions, and $z$ is the compressional direction. Because the instantaneous stress values are noisy, owing to the thermal fluctuation, the presented time-evolution data of the stress values were smoothed using the Savitzky--Golay filter~\cite{SavitzkyGolay1964}, which can smooth data without loss of the data tendency.

\begin{figure}[htbp]
    \centering
    \includegraphics[width=7in]{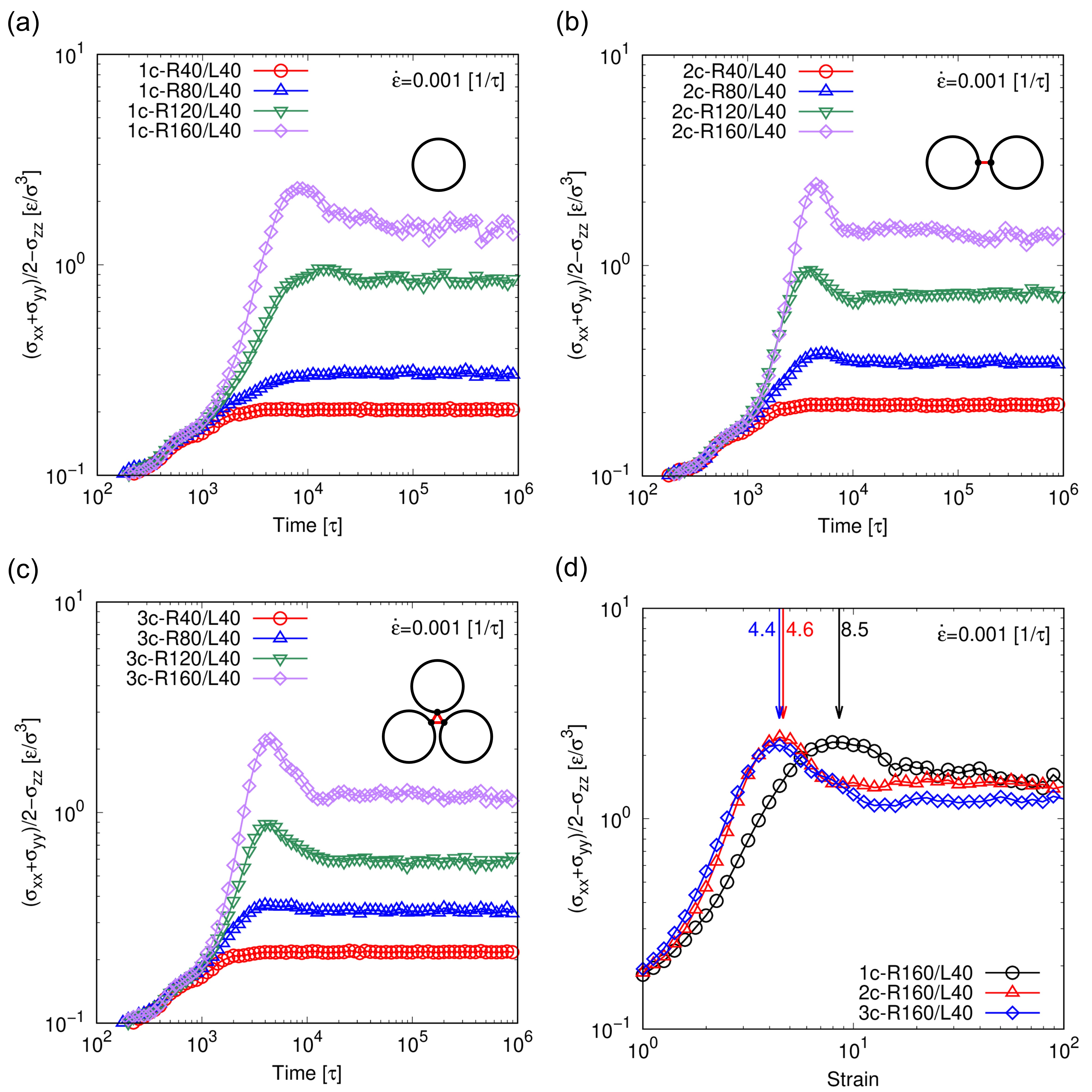}
    \caption{Time evolution of the first normal stress difference on (a) monocyclic/linear, (b) bicyclic/linear, and (c) tricyclic/linear blends under biaxial elongational flow with a strain rate of 0.001$/\tau$. The blend ratio of the cyclic/linear chains was 1/10. Blends with four different ring sizes ($N_{\rm R}$ = 40, 80, 120, and 160) were compared, while the linear chain length $N_{\rm L}$ was fixed at 40. (d) Comparison of the stress-overshoot behaviors in the three different blends with $N_{\rm R}$ = 160. For the horizontal axis in (d) the Hencky strain $\varepsilon=t\dot{\varepsilon}$.}
    \label{fig2_MonoBiTriCompare}
\end{figure}

Figures \ref{fig2_MonoBiTriCompare}a--c show the $N_{\rm R}$-dependences of stress-overshoot phenomena under biaxial elongational flow with $\dot{\varepsilon}$ = 0.001$/\tau$ for the monocyclic/linear, bicyclic/linear, and tricyclic/linear blends ($N_{\rm L}$ = 40), respectively. In the monocyclic/linear blends, stress-overshoot behavior is apparent for $N_{\rm R}$ = 160, as observed in our previous work~\cite{MurashimaEtal2021}. For $N_{\rm R}$ = 120, there is a small peak after the steep rise in the first normal stress difference, while for $N_{\rm R}$ = 40 and 80, no stress overshoot is observed. Because more than two chains penetrate a ring for $N_{\rm R}$ = 120 and 160~\cite{HagitaMurashima2021a}, the linear chains expand the ring outwardly under biaxial elongational flow. On the other hand, for $N_{\rm R}$ = 40 and 80, some rings undergo single-chain penetration while many rings display no penetration. In this situation, the force that pushes the ring outward is weak. Therefore, there is no stress-overshoot behavior for $N_{\rm R}$ = 40 and 80 in the monocyclic/linear blends.

In the multicyclic/linear blends, for $n$ = 2 and 3, we observed the stress-overshoot behaviors for $N_{\rm R}$ = 80, 120, and 160 but not for $N_{\rm R}$ = 40, reflecting whether the multicyclic chains have multiple-chain penetration at the equilibrium state~\cite{HagitaMurashima2021b}. Surprisingly, the stress overshoot observed in the multicyclic/linear blends is much steeper than that in the monocyclic/linear blends. Figure \ref{fig2_MonoBiTriCompare}d compares the stress-overshoot behaviors, at $N_{\rm R}$ = 160, of the monocyclic/linear and multicyclic/linear blends. Here, the horizontal axis is replaced with the Hencky strain ($\varepsilon = t\dot{\varepsilon}$) for convenience in later discussion. The stress-overshoot peak appears at $\varepsilon\sim$8 for the monocyclic/linear blends and $\sim$4 for the multicyclic/linear blends. The peak strain values were estimated from the least-squares fitting to a quadratic function as 8.5 for $n$ = 1, 4.6 for $n$ = 2, and 4.4 for $n$ = 3 (Figure \ref{fig2_MonoBiTriCompare}d). Although the tricyclic/linear blend presented a smaller peak strain value than that of the bicyclic/linear blend, the difference is within the statistical error margin. Moreover, Figure \ref{fig2_MonoBiTriCompare}d shows that the steady-state value of the first normal stress difference decreases with increasing $n$. These findings indicate the different origins of the stress-overshoot behavior in the monocyclic/linear and multicyclic/linear blends, the origin of which is discussed in Sections \ref{Sec3.2:Ring-shape} and \ref{Sec3.3:Topological}.

\begin{figure}[htbp]
    \centering
    \includegraphics[width=7in]{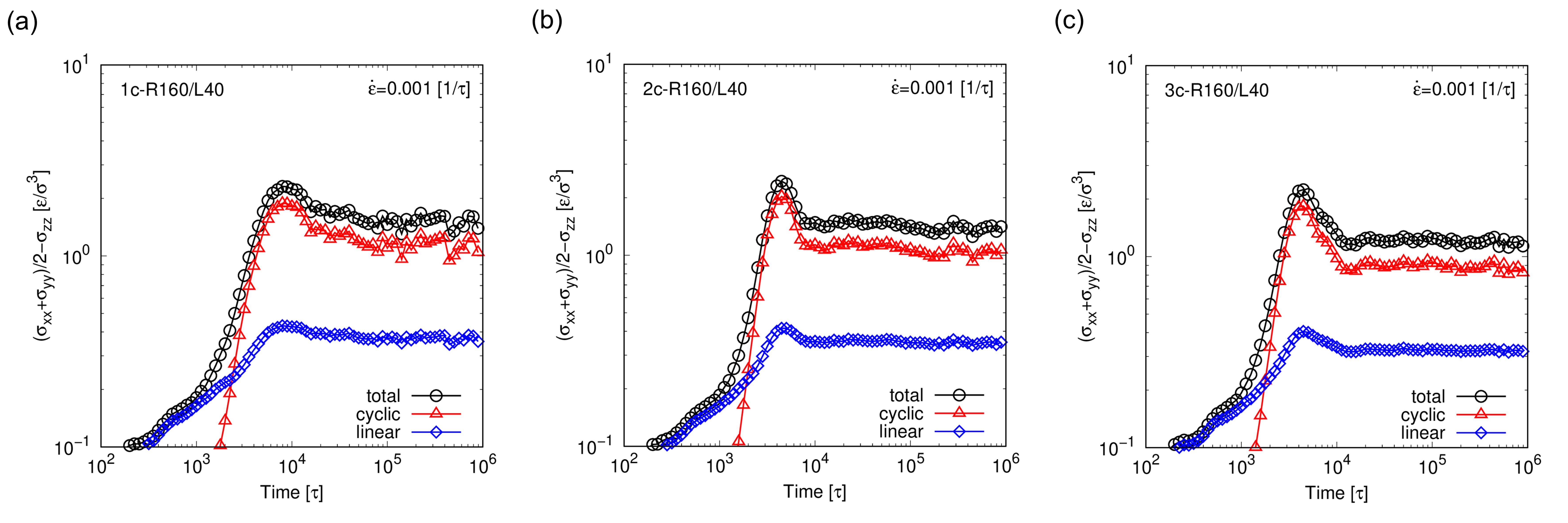}
    \caption{Decomposed stress evolution on (a) monocyclic/linear (1c-R160/L40), (b) bicyclic/linear (2c-R160/L40), and (c) tricyclic/linear (3c-R160/L40) blends under biaxial elongational flow with a strain rate of 0.001$/\tau$. The total stress is decomposed into cyclic- and linear-chain stresses.}
    \label{fig3_component}
\end{figure}

Figures \ref{fig3_component}a--c show the contributions from the cyclic and linear chains to the stress-overshoot behaviors at $N_{\rm R}$ = 160 and $N_{\rm L}$ = 40 for the monocyclic/linear, bicyclic/linear, and tricyclic/linear blends, respectively, under biaxial elongational flow with a strain rate of 0.001$/\tau$. The total stress was decomposed into the partial stresses of the cyclic and linear chains. To obtain the partial stresses, the stress on each particle was calculated, and the obtained particle stresses were summed up separately for the cyclic and linear chains. In the early stage for $t <$10$^3 \tau$, the linear-chain stress is dominant, while when the Hencky strain ($\varepsilon=t\dot{\varepsilon}$) exceeds 1.0 ($t$ = 10$^3 \tau$ in the case of $\dot{\varepsilon}$ = 0.001$/\tau$) the cyclic-chain stress exceeds the linear-chain stress. Notably, stress-overshoot behavior is observed in both the cyclic- and linear-chain stresses. The peaks for these two stresses correspond to each other and appear at $\varepsilon\sim$8 in the monocyclic/linear blend and $\sim$4 in the multicyclic/linear blends. These results indicate that as expected, the stress-overshoot behavior is caused by the cooperative dynamics between the cyclic and linear chains. 

The rapid increase for $1 < \varepsilon < 4$ in the multicyclic/linear blends originates from the difference in the end-to-end distance between the monocyclic and multicyclic chains, where the end-to-end distance of a ring is defined as the distance between two particles separated by $N_{\rm R}/2 - 1$ bonds along the chain. The end-to-end distances of the multicyclic chains for $n$ = 2 and 3 are defined by two end particles on two different rings in a multicyclic chain. Here, an end particle on a ring in a multicyclic chain is a particle separated from a bond-connected particle by $N_{\rm R}/2 - 1$ bonds along the chain (Figure \ref{fig1_schematics}). When the rings expand under biaxial elongational flow, the end-to-end distance of the multicyclic chain is longer than that of the monocyclic chain, owing to the extra ring. 
In addition, the multicyclic chains have a branch point, while the monocyclic chain does not. In analogy with the difference between stars and linear chains, the branch point slows down the dynamics of multicyclic chains~\cite{DoiEdwards1986,MilnerMcLeish1997,HuangEtal2016}. Because of the longer end-to-end distance and the presence of a branch point in the multicyclic chains, the relaxation time is longer in the multicyclic chains than in the monocyclic chain. Thus, at the same strain rate, the multicyclic chains have a higher Rouse Weissenberg number ($Wi_{\rm R}=\dot{\varepsilon}\tau_{\rm R}$, where $\tau_{\rm R}$ is the Rouse relaxation time) than that of the monocyclic chain. The Rouse Weissenberg number indicates the strength of nonlinearity under elongational flow, and thus, strain hardening is steeper in the multicyclic/linear blends than in the monocyclic/linear blend.

The reason why the overshoot peak appears earlier for the multicyclic/linear blends than for the monocyclic/linear blends remains unclear. However, a possible reason was revealed by molecular shape analyses, as discussed later. Notably, the slopes of the reduction process are near-identical for the bicyclic/linear and tricyclic/linear blends (Figure \ref{fig2_MonoBiTriCompare}d); however, the steady state appears later for the latter blend. This implies that the bicyclic/linear and tricyclic/linear blends share the same reduction mechanism; however, the reduction time is longer for the latter blend. This further reduction observed in the tricyclic linear blend is also discussed in Section \ref{Sec3.3:Topological}.

\begin{figure}[htbp]
    \centering
    \includegraphics[width=7in]{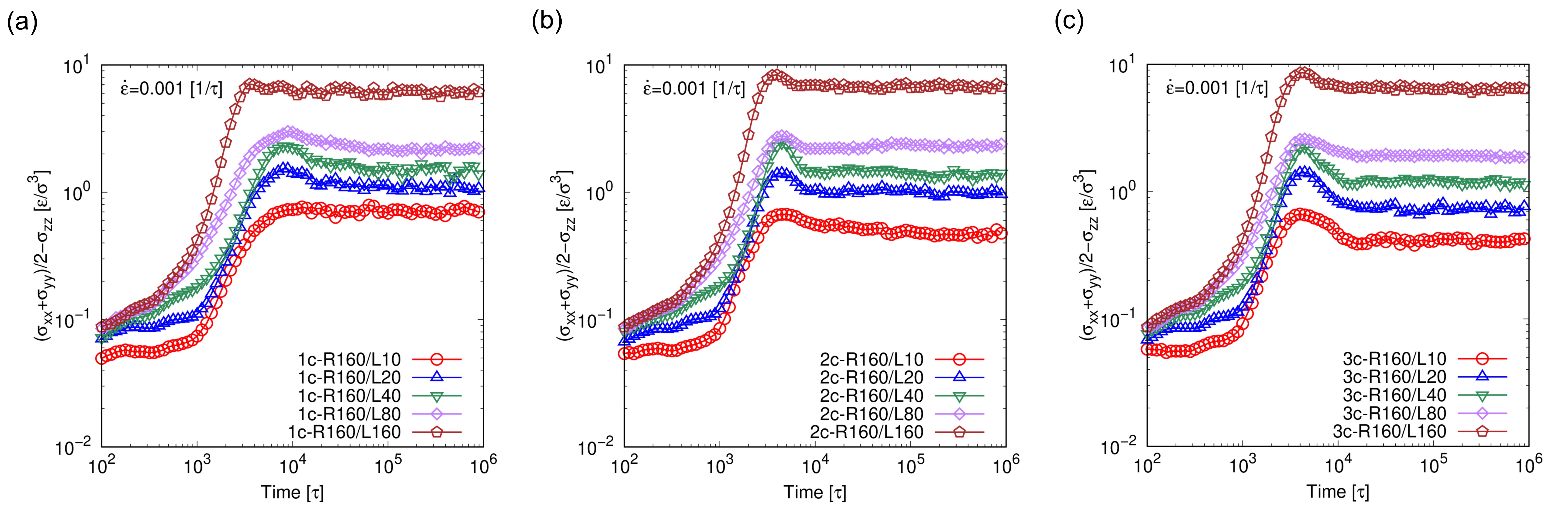}
    \caption{$N_{\rm L}$-dependence of the stress evolution in the (a) monocyclic/linear, (b) bicyclic/linear, and (c) tricyclic/linear blends under biaxial elongational flow with a strain rate of 0.001$/\tau$. Five different linear chain lengths ($N_{\rm L}$ = 10, 20, 40, 80, and 160) were compared, while the ring size $N_{\rm R}$ was fixed at 160.}
    \label{fig4_NLdependense}
\end{figure}

Figures \ref{fig4_NLdependense}a--c show the $N_{\rm L}$-dependences of the time-evolution behaviors of the first normal stress difference under biaxial elongational flow with $\dot{\varepsilon}$ = 0.001$/\tau$ for the monocyclic/linear, bicyclic/linear, and tricyclic/linear blends ($N_{\rm R}$ = 160), respectively. Although the data of the monocyclic/linear blends shown in Figure \ref{fig4_NLdependense}a are the same as those presented in our previous work~\cite{MurashimaEtal2021}, we replotted them for comparison with the data of the multicyclic/linear blends. The data revealed that the $N_{\rm L}$-dependencies of the monocyclic/linear and multicyclic/linear blends are significantly different. Specifically, the stress overshoot is only significant at $N_{\rm L}$ = 40 in the monocyclic/linear blends, whereas sharp stress overshoot is observed for all $N_{\rm L}$ values in the multicyclic/linear blends.

In the monocyclic/linear blends, strain hardening (nonlinear increase for $\varepsilon >$1; $t >$10$^3 \tau$ for $\dot{\varepsilon}$ = 0.001$/\tau$) occurs in both the long and short chains. The rings expand under biaxial elongational flow with $\dot{\varepsilon}$ = 0.001$/\tau$, regardless of the $N_{\rm L}$ value. The overshoot when $N_{\rm L}$ = 40 affords the most notable behavior because the ring bond lengths, which are stretched rather than in equilibrium, are maximized when $N_{\rm L}$ = 40. These findings, which are reported in our previous work~\cite{MurashimaEtal2021}, occur because the short linear chains ($N_{\rm L} <$20) are not sufficient to extend the bonds of the rings, whereas the long linear chains ($N_{\rm L} >$80) disturb the extra ring extension, owing to the entanglement network consisting of linear chains.

In contrast, the multicyclic linear blends exhibited stress-overshoot behavior, regardless of the $N_{\rm L}$ value. Comparison of the bicyclic/linear and tricyclic/linear blends revealed that the overshoot becomes more significant as $n$ increases. Because in multicyclic/linear blends this overshoot is observed even in short linear chains, we deduced that the overshoot is not caused by bond elongation on the rings. Moreover, this overshoot is not affected by the entanglements of the long linear chains. Thus, evidently, the overshoot mechanism in multicyclic/linear blends is different from that in monocyclic/linear blends, which origin is the extra ring elongation followed by slight ring shrinkage. Notably, the overshoot displays $N_{\rm R}$-dependency, as shown in Figure \ref{fig2_MonoBiTriCompare}. The overshoot in multicyclic/linear blends is caused by the linear-chain penetration, in particular multiple-chain penetrations, similar to those observed in the monocyclic/linear blends.

The above results revealed that the overshoot behaviors in multicyclic/linear blends are phenomena in which multiple linear chains penetrate and expand the ring; however, overshoot occurs without extra bond elongation and shrinkage of the ring. It is considered that the origin of multicyclic/linear blends is different from that of monocyclic/linear blends. In the case of multicyclic/linear blends, we established that molecular morphology analyses should be performed to investigate the cause of the overshoot behavior in more detail. The next section focuses on bicyclic/linear blends for morphological analysis.

\clearpage

\subsection{Ring-shape analysis on bicyclic chains under biaxial elongational flow}\label{Sec3.2:Ring-shape}
\begin{figure}[htbp]
    \centering
    \includegraphics[width=7in]{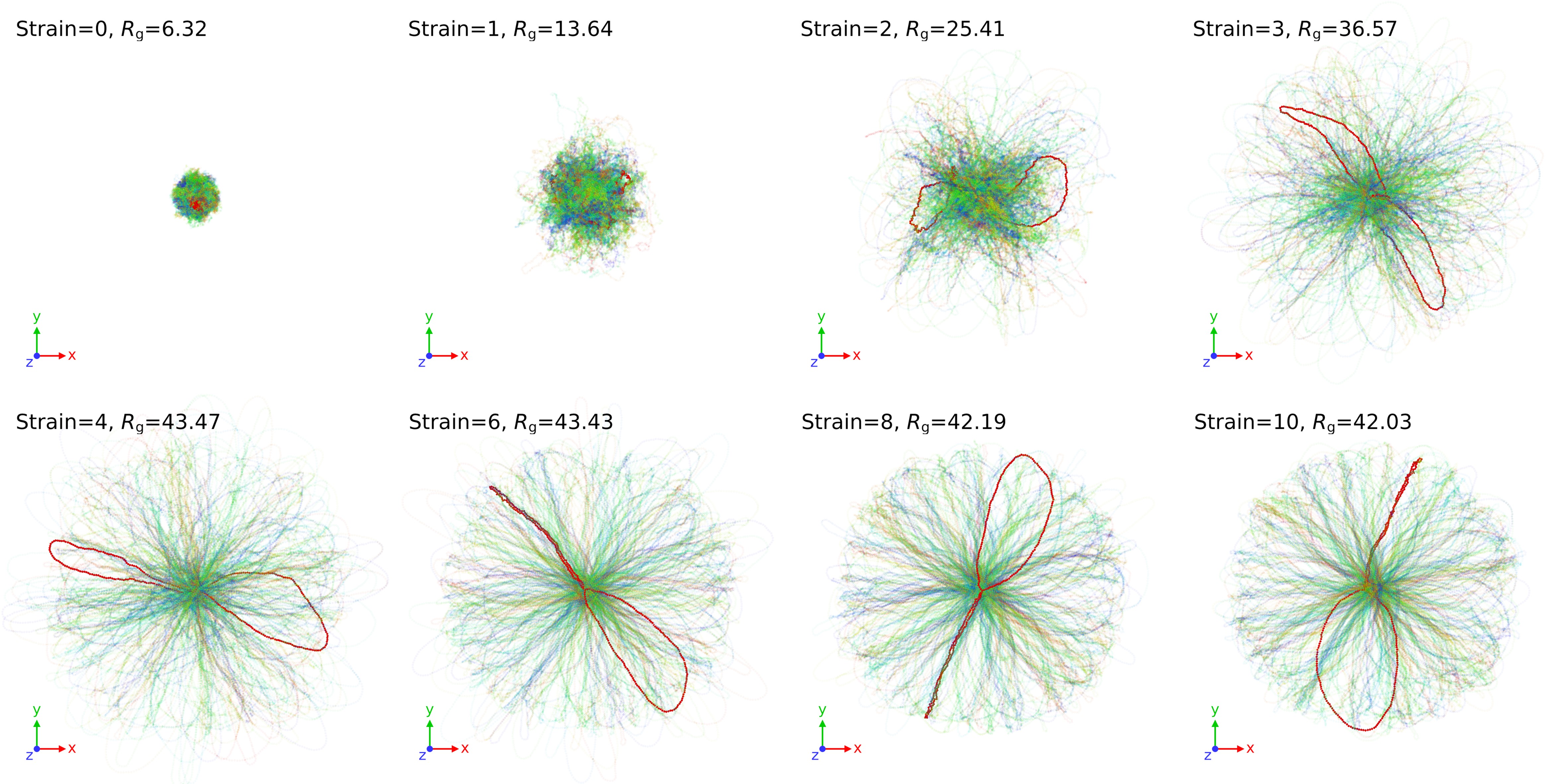}
    \caption{Strain-dependent mean shape of the bicyclic chains in the bicyclic/linear blends (2c-R160/L40) under biaxial elongational flow with a strain rate of 0.001$/\tau$. The bicyclic chains are presented and viewed from the $z$-axis (compression direction). The mass center of each bicyclic chain is shifted to the origin (0,0,0), and the average value of the gyration radius $R_{\rm g}$ of the bicyclic chains is presented with the strain value. The red thick bicyclic chain in each snapshot has the closest value of $R_{\rm g}$ to its average value at each strain.}
    \label{fig5_averageshape}
\end{figure}

This section discusses the cause of overshoot arising at a strain $\varepsilon$ of $\sim$4 in multicyclic/linear blends through molecular morphology analysis. We first focus on the behavior of bicyclic chains in a bicyclic/linear blend, with $N_{\rm R}$ = 160 and $N_{\rm L}$ = 40, under biaxial elongational flow with $\dot{\varepsilon}$ = 0.001$/\tau$. Figure \ref{fig5_averageshape} shows superimposed snapshots of the bicyclic chains at each strain, which were captured by moving the bicyclic chains in the system so that their mass centers overlapped at the origin of the coordinates. The value of the gyration radius $R_{\rm g}$ of the bicyclic chains, presented in each snapshot, is an average value of all the bicyclic chains in the system. Each molecule is color coded according to its identification number. The red thick molecule displayed the closest $R_{\rm g}$ value to that of the mean shape in the system. Thus, we concluded that the shape of the red thick molecule best approaches the mean shape. When the strain is equal to zero, the mean shape of the bicyclic chain is the equilibrium coiled state. As the strain increases, the coiled state of the bicyclic chain is compressed in the $z$ direction and radially spreads in the $xy$ plane. The two rings open and broaden up to $\varepsilon$ = 4. Surprisingly, the mean shapes at $\varepsilon$ = 6, 8, and 10 represented a tadpole-like shape of the bicyclic chain, where the hole in one of the rings is tightly closed to form a double-folded string. On the other hand, for the monocyclic chains, this hole remained open under biaxial elongational flow, and such a closed ring was not observed in our previous work (see Figure 6 in Ref. \citenum{MurashimaEtal2021}). Furthermore, because the morphological (open-to-closed) transition of the ring occurs when the strain in the bicyclic chains is $>$4, we supposed that this transition is strongly related to the overshoot behavior.

To observe in detail the changes in the ring morphology, we next introduced two ring internal lengths, $l_1$ and $l_2$. Ring internal length $l_1$ is the distance between a bond-connected (first) particle and an end (($N_{\rm R}$/2)-th) particle on the same ring, separated by $N_{\rm R}/2 - 1$ bonds along the ring from the bond-connected particle, while $l_2$ is the distance between the ($N_{\rm R}$/4)-th and (3$N_{\rm R}$/4)-th particles. Notably, the gyration tensor is generally used to characterize the morphology of a molecule. However, attempts to distinguish between the open and closed states using the gyration tensor failed. This is because the gyration tensor approximates a molecule as an ellipsoidal shape, and the eigenvalues of the gyration tensor are therefore insensitive to the open and closed states of the rings. Thus, we used $l_1$ and $l_2$ to capture the open and closed states under biaxial elongational flow.

\begin{figure}[htbp]
    \centering
    \includegraphics[width=7in]{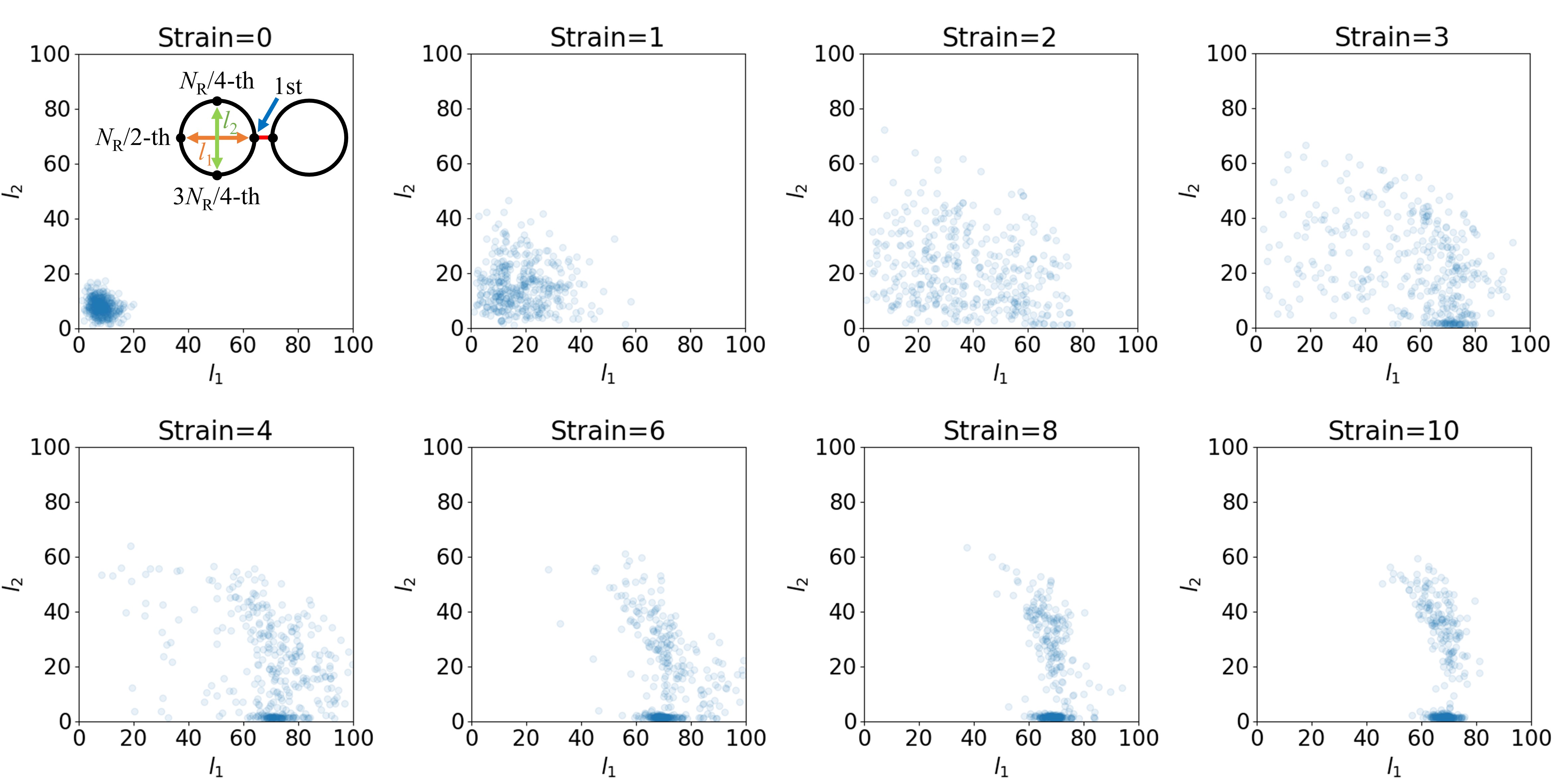}
    \caption{Strain-dependent scatter plots of $l_1$ versus $l_2$ in the bicyclic/linear blend (2c-R160/L40) under biaxial elongational flow with a strain rate of 0.001$/\tau$. Ring internal length $l_1$ is the distance between a bond-connected (first) particle and an end (($N_{\rm R}$/2)-th) particle on the same ring, separated by $N_{\rm R}/2 - 1$ bonds along the ring from the bond-connected particle, while $l_2$ is the distance between the ($N_{\rm R}$/4)-th and (3$N_{\rm R}$/4)-th particles.}
    \label{fig6_2dmap}
\end{figure}

Figure \ref{fig6_2dmap} shows scatter plots of $l_1$ versus $l_2$ for each strain: When $\varepsilon$ = 0, the points are distributed at approximately $l_1$ = $l_2$ = 10. As the strain increases ($\varepsilon <$2), the points are widely dispersed, while at $\varepsilon$ = 3, a concentrated region at $60 < l_1 < 80$ and $l_2 <$3 is observed. With the further increase in the strain ($\varepsilon >$4), the distribution is divided into two regions, namely a dense region ($60 < l_1 < 80$ and $l_2 <$3) comprising closed rings and a low-density region, in which the open rings are widely distributed. Because the dense region contains rings with a tightly constricted center, $l_1$ and $l_2$ were not sufficient to determine the closed-ring state. Thus, to determine the closed-ring state more precisely, we introduced auxiliary ring internal lengths, $l_3$ and $l_4$, where $l_3$ is the distance between the ($N_{\rm R}$/8)-th and (7$N_{\rm R}$/8)-th particles and $l_4$ is the distance between the (3$N_{\rm R}$/8)-th and (5$N_{\rm R}$/8)-th particles.

\begin{figure}[htbp]
    \centering
    \includegraphics[width=7in]{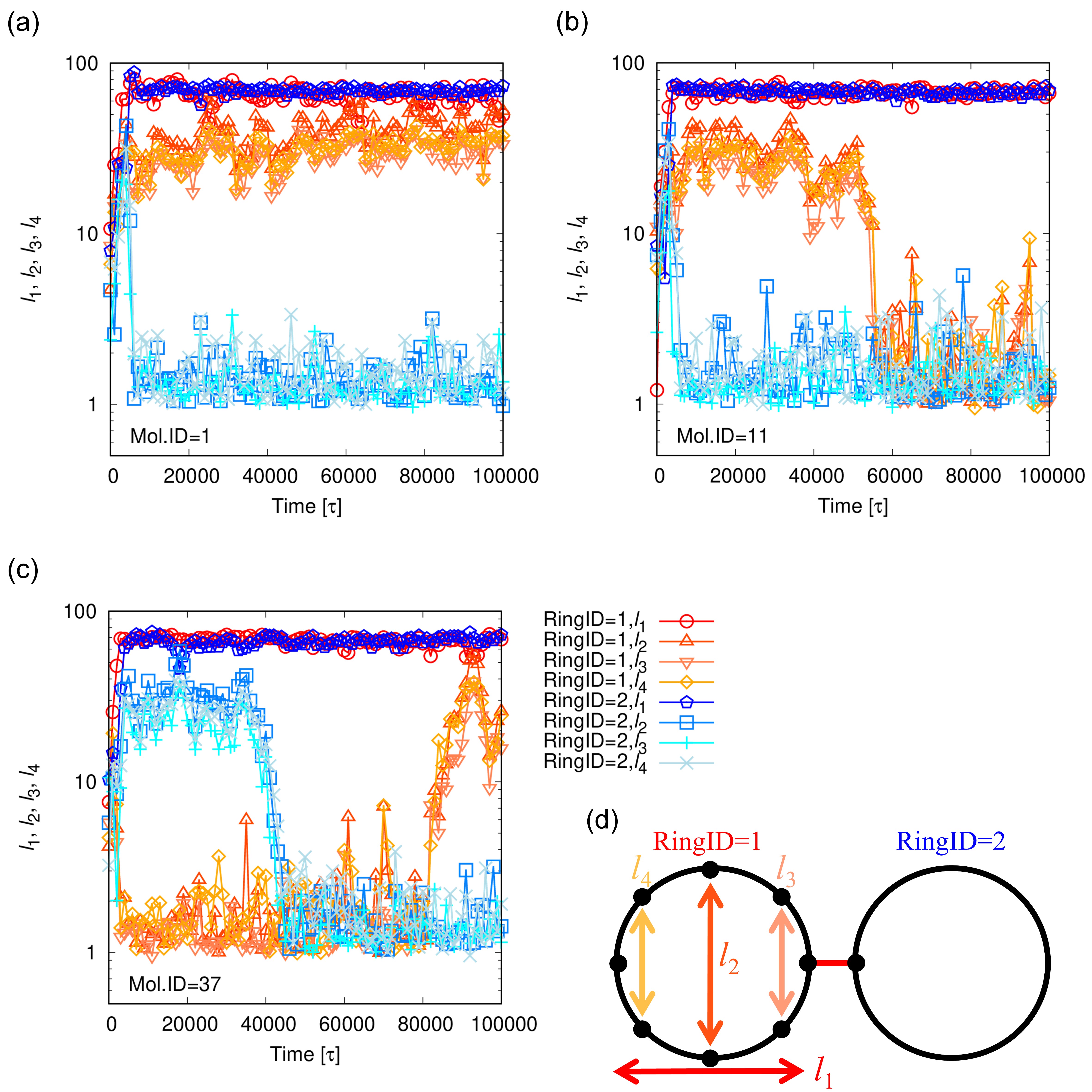}
    \caption{Time-evolution behaviors of four ring internal lengths ($l_1$--$l_4$) for two rings on molecules with Mol.ID = 1 (a), 11 (b), and 37 (c) in the bicyclic/linear blend (2c-R160/L40) under biaxial elongational flow with a strain rate of 0.001$/\tau$. (d) Schematic explanations of $l_1$--$l_4$. The Mol.ID is the molecule identification number, while the RingID represents one of the two rings in a bicyclic chain.}
    \label{fig7_transition}
\end{figure}

Figure \ref{fig6_2dmap} reveals the presence of two types of ring shapes, namely open- and closed-ring states. To characterize the morphology of the bicyclic chains under biaxial elongational flow, we analyzed four ring internal lengths ($l_1$--$l_4$) for each ring in the bicyclic/linear blend (2c-R160/L40) under biaxial elongational flow. The time-evolution behaviors of $l_1$--$l_4$ for molecule identification numbers (Mol.IDs) 1, 11, and 37 are presented in Figures \ref{fig7_transition}a--c, respectively, wherein the RingID represents one of the two rings in a bicyclic chain. In addition, the schematic explanation of $l_1$--$l_4$ is presented in Figure \ref{fig7_transition}d. The $l_1$ values for both rings monotonically increase for $0 < t < 10,000$ and then represent steady states for $t >$10,000. In Figures \ref{fig7_transition}a--c, the $l_1$ behaviors are similar in all three graphs, while at the steady state of $l_1$ ($t >$10,000) the behaviors of $l_2$--$l_4$ in the three graphs exhibit some differences. Figure \ref{fig7_transition}a shows the most typical case in which, in the steady state, one of the rings has small $l_2$--$l_4$ values indicating a slightly open ring, while the other ring has large $l_2$--$l_4$ values indicating a fully open ring. A total of 136 out of 192 samples exhibited similar behaviors. For $t >$10,000, the respective estimated means and standard deviations are 1.43 and 0.39 for $l_2$, 1.35 and 0.39 for $l_3$, and 1.57 and 0.5 for $l_4$ on the slightly open ring and 40.6 and 8.9 for $l_2$, 29.5 and 5.6 for $l_3$, and 31.8 and 5.3 for $l_4$ on the fully open ring. Although the values of $l_2$--$l_4$ for the slightly open ring sometimes reached a value of 3, as shown in Figure \ref{fig7_transition}a, the ring tended to be closed. This implies that the slightly open ring was stable at $l_2$--$l_4$ values $<$3. A transition between the fully and slightly open rings was sometimes observed in the steady state, as shown in Figure \ref{fig7_transition}b. Moreover, a rare example of alternating the fully and slightly open rings is shown in Figure \ref{fig7_transition}c. From these transitions, we could roughly estimate the lifetime of the steady-state morphology of the bicyclic chain under biaxial elongational flow. When counting the number of transitions, we ignored any instantaneously large fluctuations, which can be interpreted as precursory phenomena before the transition. Thus, we determined that one transition occurred when the values of $l_2$--$l_4$ changed from $<$3 to $>$15 (or vice versa) and then stabilized for several time steps. Values of 3 and 15 were selected by three standard deviations from the mean value of $l_4$ for the slightly open ring and fully open ring, respectively. Figure \ref{fig7_transition}b illustrates one transition, while in Figure \ref{fig7_transition}c the transition occurs twice. For a total of 192 samples in the current system, the transition was observed once in 38 samples and twice in 18 samples, while 136 samples presented no transitions in the range $10,000 < t < 100,000$. Thus, the probability of a transition being observed at $10,000 < t < 100,000$ was approximately 0.39. The lifetime was approximately 230,000$\tau$ ($\approx$ 90,000/0.39), which is sufficiently long under biaxial elongational flow with $\dot{\varepsilon}$ = 0.001$/\tau$ (Hencky strain = 230). Hereafter, a ring is defined as being in the closed-ring state when it satisfies the conditions $l_1 >$60, $l_2 <$3, $l_3 <$3, and $l_4 <$3, labeled as the closed-ring condition; otherwise, the ring is considered to be in an open-ring state. According to this definition, the rings in equilibrium are the open-ring state. A slightly open ring with a large $l_1$ is regarded as the closed-ring state, accounting for a ring transiently opened by thermal fluctuations. Notably, the condition $l_1 >$60 depends on $N_{\rm R}$, whereas the other conditions are independent of $N_{\rm R}$.

\begin{figure}[htbp]
    \centering
    \includegraphics[width=7in]{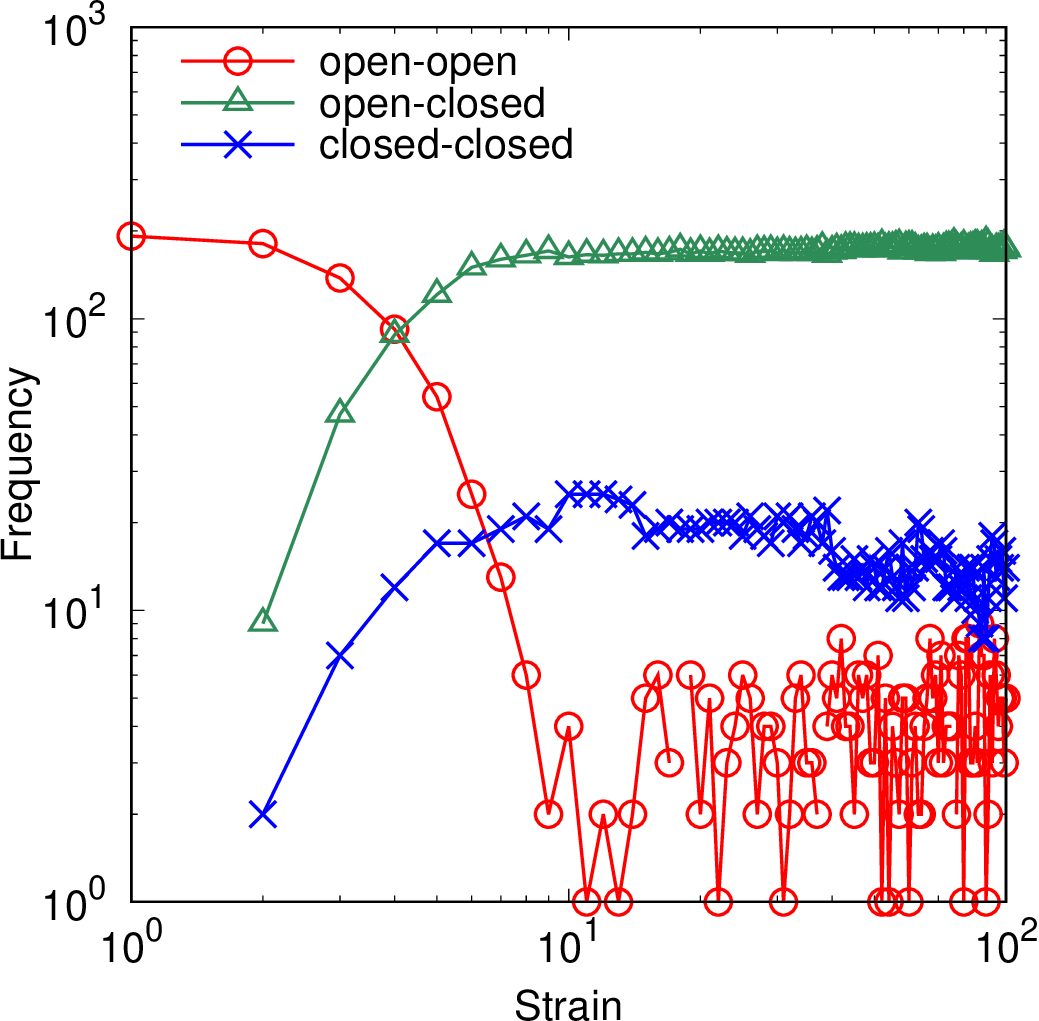}
    \caption{Appearance frequency versus strain in the bicyclic/linear blend (2c-R160/L40) under biaxial elongational flow with a strain rate of 0.001$/\tau$. Three typical states are observed in the bicyclic chains: open-open, open-closed, and closed-closed states.}
    \label{fig8_shapeinfo}
\end{figure}

In the case of bicyclic chains, three combinations (open-open, open-closed, and closed-closed) were considered. Examining the strain dependency of the frequency of these three combinations revealed some interesting facts, as shown in Figure \ref{fig8_shapeinfo}. As the strain increases, the number of open-open states decreases, whereas those of open-closed and closed-closed states increase. At $\varepsilon$ = 4, the magnitude relationship between the open-open and open-closed is reversed. Regarding the contribution to the first normal stress difference under biaxial elongational flow, that of the open ring is higher than that of the closed ring. This is because the former has the force to push the ring outward, owing to the linear chains penetrating it, whereas the closed ring does not. Thus, the morphological change of the ring causes stress overshoot.

\begin{figure*}[htbp]
    \centering
    \includegraphics[width=7in]{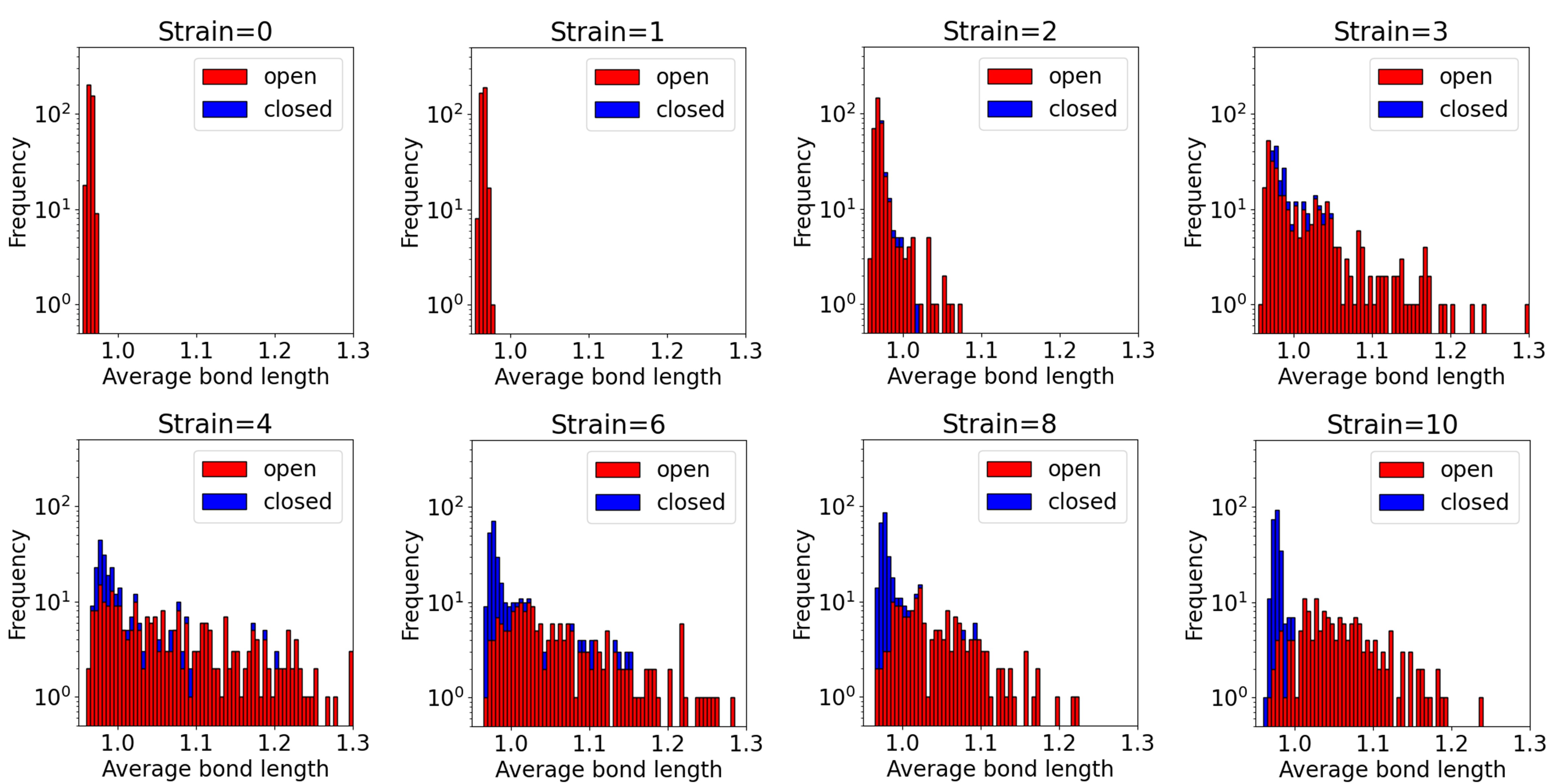}
    \caption{Strain-dependent distribution of the average bond lengths on the open and closed states of the rings in the bicyclic/linear blends (2c-R160/L40) under biaxial elongational flow with a strain rate of 0.001$/\tau$. The bond lengths are averaged for one ring.}
    \label{fig9_histbondlength}
\end{figure*}

Figure \ref{fig9_histbondlength} shows the relationship between the state of the ring and its bond length for each strain; the bond lengths are averaged for each ring. At the initial state ($\varepsilon$ = 0), all rings are in the open state, and the average bond length in each ring is $<$1.0. This is because the equilibrium bond length of the KG model is 0.97. As the strain increases, the average bond length also increases. Moreover, at $\varepsilon$ = 2, closed states begin to appear, until for $\varepsilon >$4, the closed states are distributed within the range where the average bond length is $<$1.0. This implies that the bond in the closed state is in a relaxed state. However, many bonds in the open state have a bond length that is much larger than the equilibrium length. Therefore, we supposed that the morphological transition, from the open to closed state, causes bond relaxation and a decrease in tension at the joint connecting two rings.

The reason why the closed state does not occur in a single ring but in a bicycle can be understood from the morphological difference between the two. In the former, the center of the molecule is the void of the ring, whereas in the latter, the center is the joint connecting the two rings. Because the molecules are expanded in the $xy$ plane under biaxial elongational flow, single rings are isotropically expanded, whereas in a bicyclic chain, the individual rings spread radially around the joint connecting two rings. In addition, under elongational flow, the farther the molecule is from its center, the greater is the deformation. Therefore, in the case of a bicyclic chain, the holes of both rings tend to close, whereas the linear chains penetrating the rings disturb ring closure. When the linear chains penetrating one of the rings disappear, the ring without linear chain penetration approaches a double-folded string shape, which is the characteristic (a non-obstacle-enclosing and non-ramified) shape at equilibrium~\cite{Klein1986}. A double-folded string shape has also been considered in nonlinear shear flow in the shear slit model~\cite{ParisiEtal2021a,ParisiEtal2021b}. Under biaxial elongational flow, the double-folded string is confined by the surrounding chains and is highly stretched, whereas the orientation is random in the elongational plane. Once linear chains penetrating a ring are pulled out, the ring loses the tension spread outward by the penetration of the linear chain, and then forms a double-folded string shape. If either of the two rings forms a double-folded string first, the balance between them is broken, and the joint relaxes. As the tension at the joint weakens, the remaining ring behaves as a single ring and stabilizes in an expanded state. From the above considerations, we concluded that a morphological transition causes the stress-overshoot behavior of the multicycle/linear blends, and the open-closed state in a bicycle is dominant under biaxial elongational flow with $\dot{\varepsilon}$ = 0.001$/\tau$.

The morphological transition in bicyclic chains can be mathematically explained in terms of the topology. Thus, a bicyclic chain with two open rings (open-open state) is regarded as genus 2. Similarly, the open-closed state is regarded as genus 1, and the closed-closed state as genus 0. Specifically, the number of open rings in a multicyclic chain is regarded as the value of its genus. This is not just wordplay but also a physically meaningful interpretation. As previously discussed, an open ring allows chain penetration, whereas a closed ring inhibits it. The open ring is pushed outward by the chains penetrating it, whereas the closed ring does not experience such a force. This implies that the topological transition from the open- to closed-ring state drastically changes the stress of the ring. It is therefore better to consider stress-overshoot behavior as a difference in the topology rather than simply a morphological change. Notably, the maximum value of the genus in a bicyclic chain is 2. Although an open ring can have several holes depending on its morphology, the value of the genus was evaluated as 1 in this study. Therefore, a tricyclic chain can assume a state ranging from genus 0 to genus 3. To further understand the topological transition, tricyclic chains are analyzed in the next section. 

\clearpage

\subsection{Topological transition in a tricyclic/linear blend under biaxial elongational flow}\label{Sec3.3:Topological}
\begin{figure}[htbp]
    \centering
    \includegraphics[width=7in]{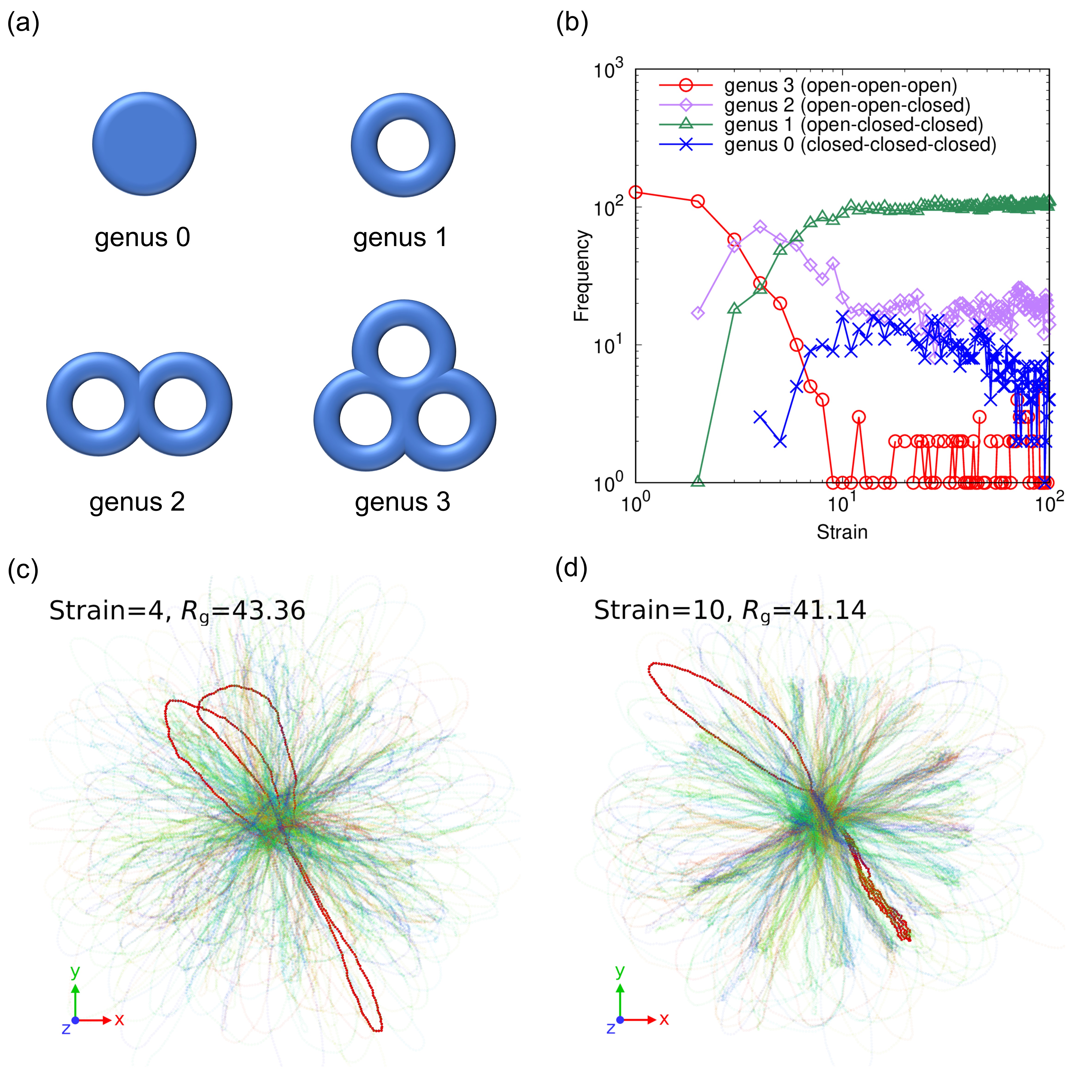}
    \caption{(a) Schematic images of the four genera. (b) Appearance frequency versus strain in the tricyclic/linear blend (3c-R160/L40) under biaxial elongational flow with a strain rate of 0.001$/\tau$. Four typical states are classified by four genera: genus 3 (open-open-open), genus 2 (open-open-closed), genus 1 (open-closed-closed), and genus 0 (closed-closed-closed). (c, d) Mean shapes of the tricyclic chains at $\varepsilon$ = 4 and 10.}
    \label{fig10_tricyclic}
\end{figure}

As shown in the previous section, the morphological changes in bicyclic chains induced stress-overshoot behavior. This phenomenon can be explained from the topology of the bicyclic chains. Thus, in this section, we analyze tricyclic/linear blends under biaxial elongational flows in terms of topology. A tricyclic chain has four topologically different states: open-open-open, open-open-closed, open-closed-closed, and closed-closed-closed. These four states can be classified by their genus. In this study, we regarded the number of open rings in a multicyclic chain as the value of the genus, and thus, a tricyclic chain is expected to represent four genera ranging from 0 to 3. Schematic images of the four genera are shown in Figure \ref{fig10_tricyclic}a.

We analyzed the morphologies of the tricyclic chains in the tricyclic/linear blend 3c-R160/L40, under biaxial elongational flow with $\dot{\varepsilon}$ = 0.001$/\tau$. 
Based on the analyses in the previous section, the closed state of the ring is specified by the closed-ring condition $l_1 >$60, $l_2 <$3, $l_3 <$3, and $l_4 <$3. 
Figure \ref{fig10_tricyclic}b shows the frequency of the four topologies versus strain in the tricyclic/linear blend under biaxial elongational flow. 
Similar to that in Figure \ref{fig8_shapeinfo}, the genus 1 state displays the highest percentage in the steady state ($\varepsilon >$10). 
This is because the single open state is stabilized under biaxial elongational flow, as previously discussed. 
Around the stress-overshoot peak ($\varepsilon$ = 4), the genus 2 state is the most populous. Reflecting this, the mean shape of the tricyclic chains at $\varepsilon$ = 4 (Figure \ref{fig10_tricyclic}c) represents the genus 3 state; however, with one of the rings closing. 
Because the state around $\varepsilon$ = 4 is transient, it is probable that a state in the middle of the transition from genus 3 to genus 2 was observed in this case. 
For the tricyclic chain, the genus 1 state is again a tadpole-like in shape, with the two closed rings tightly stuck together, as shown in Figure \ref{fig10_tricyclic}d. Further reduction in the first normal stress difference in Figure \ref{fig2_MonoBiTriCompare}d is also understood from the topology. Both the bicyclic and tricyclic chains prefer the steady state of genus 1 under biaxial elongational flow. Specifically, the bicyclic chain closes one ring, whereas the tricyclic chain closes two rings in the steady state. Therefore, the stress reduction in the tricyclic/linear blend by the open-to-closed transition is twice that of the bicyclic/linear blend.

\begin{figure}[htbp]
    \centering
    \includegraphics[width=7in]{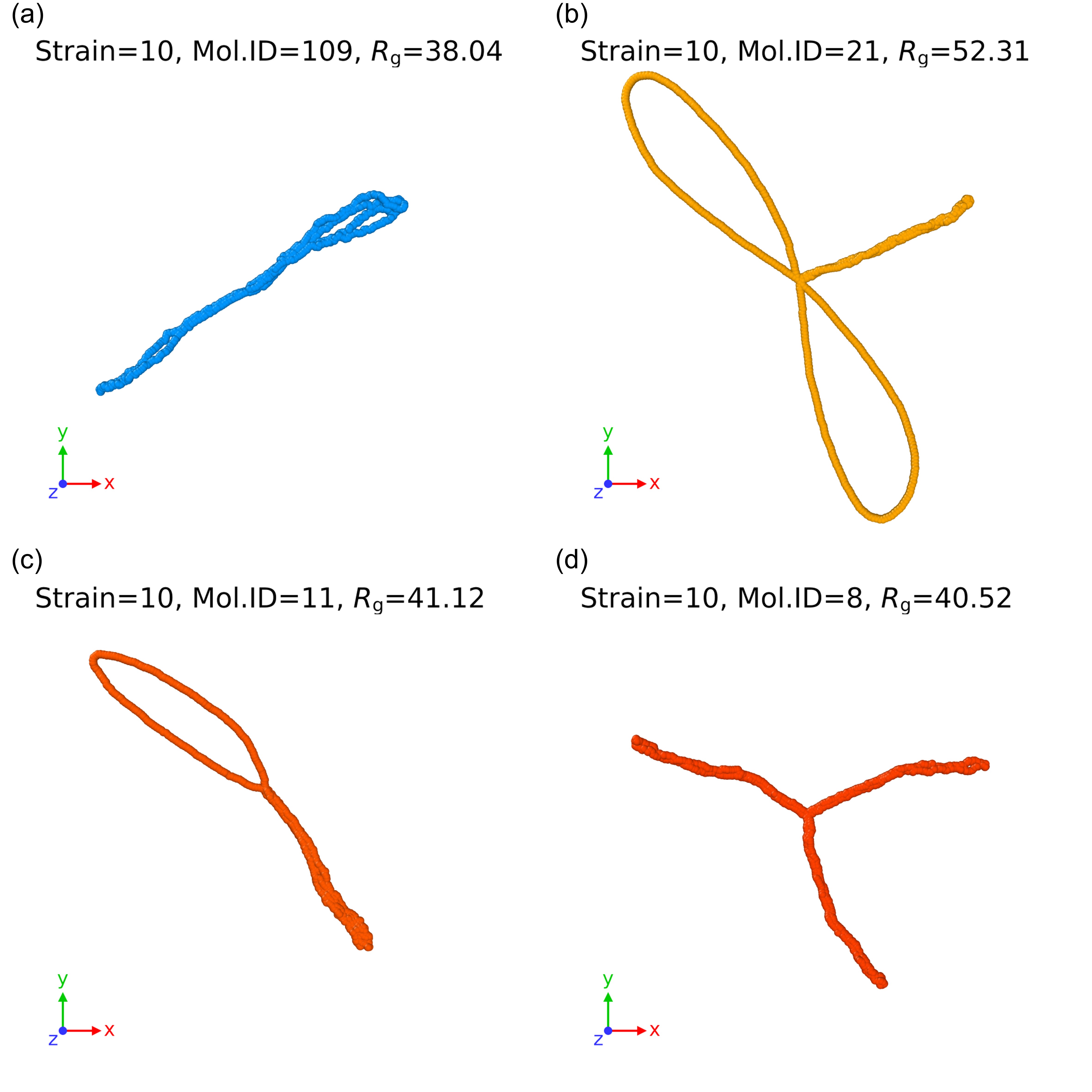}
    \caption{Four typical topological states of tricyclic chains at $\varepsilon$ = 10 in the tricyclic/linear blend (3c-R160/L40) under biaxial elongational flow with $\dot{\varepsilon}$ = 0.001$/\tau$: (a) genus 3, (b) genus 2, (c) genus 1, and (d) genus 0. Molecules are color coded according to the molecule identification number, ranging from 1 to 128. The gyration radius $R_{\rm g}$ of each molecule is also presented. The particle diameter was set to 3, accounting for the closed-ring condition for $l_2$ to $l_4$ ($l_2 <$3, $l_3 <$3, and $l_4 <$3).}
    \label{fig11_topology}
\end{figure}

Furthermore, the typical examples of the topological states on the tricyclic chains (genera 3 to 0) at $\varepsilon$ = 10 are summarized in Figures \ref{fig11_topology}a--d, respectively. 
The diameter of the particles increased up to a value of 3, corresponding to the closed-ring condition for $l_2$--$l_4$ ($l_2 <$3, $l_3 <$3, and $l_4 <$3). The figure for genus 1 in (c) corresponds to the red thick molecule shown in Figure \ref{fig10_tricyclic}d. Notably, a double-tail tadpole shape, in which two closed rings are unattached, is also observed in genus 1, although the tadpole shape shown in (c) is the most populous. Topological classification was successful, except for genus 3, where at $\varepsilon$ = 10, only one case was classified as belonging to genus 3. Although this case is very close to genus 0, the three rings break the closed-ring condition $l_2 >$3, as evidenced by the small gap in each ring in Figure \ref{fig11_topology}a. Genus 3 was considered unstable at $\varepsilon$ = 10. Genera 2 and 1, clearly comprise opened and tightly closed rings. 
The genus 2 state is a dragonfly-like in shape, while the genus 0 state is in the shape of a three-arm star. 
As shown in Figure \ref{fig10_tricyclic}b, the populations of these topologies in the steady state are of the order genus 1 $>$ genus 2 $>$ genus 0 $>$ genus 3. 
Although the single open-ring state (genus 1) is the most stable under biaxial elongational flow, as previously discussed, the state of genus 2 accounts for approximately 10\% at the large strain region. 
It should be reminded that the state of genus 2 was suppressed in the bicyclic/linear blend. This indicates that the closure of one ring contributes to the stabilization of genus 2. 
If multicyclic chains, in which more than three rings are connected, were to be considered, a stable genus 3 state might also be observed under biaxial elongational flow. 
To verify this speculation, however, further investigations are necessary.

The present study found that stress overshoot of multicyclic/linear blends under biaxial elongational flow is induced by the topological transition in symmetric multicyclic chains; however, whether asymmetric multicyclic chains exhibit stress overshoot remains an open query. Thus, we performed additional computations on asymmetric-tricyclic chains and tricyclic chains with two symmetric rings, presented in Sections \ref{Sec3.4:Asymmetric} and \ref{Sec3.5:TwoSymmetricRing}, respectively. These additional studies clarify that the symmetry of rings in a multicyclic chain plays a key role to produce stress overshoot under biaxial elongational flow.

\clearpage

\subsection{Asymmetric tricyclic chains}\label{Sec3.4:Asymmetric}
\begin{figure}[htbp]
    \centering
    \includegraphics[width=7in]{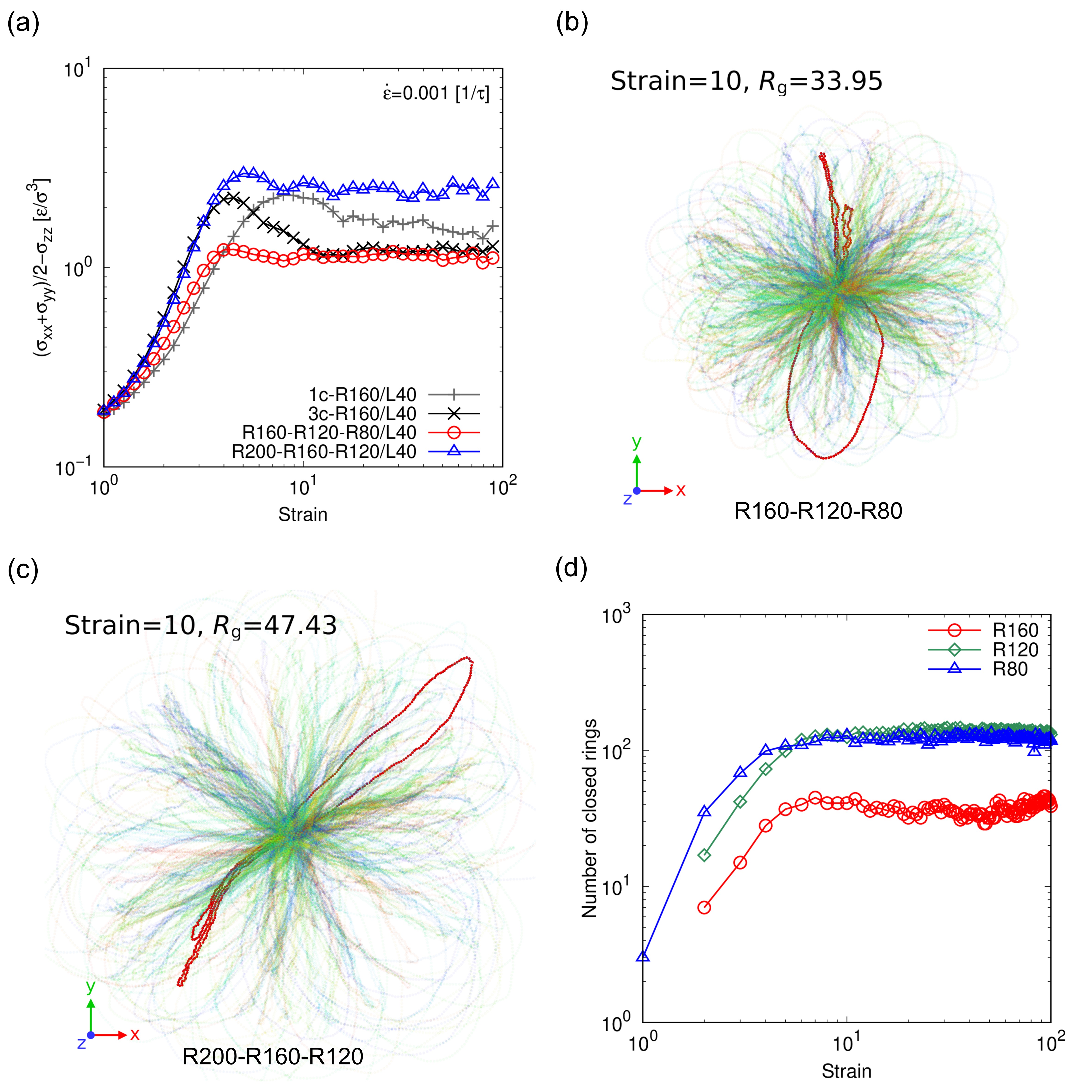}
    \caption{(a) Comparison of the stress--strain curves of R160-R120-R80/L40 and R200-R160-R120/L40, under biaxial elongational flow with a strain rate of 0.001$/\tau$, with those of 1c-R160/L40 and 3c-R160/L40. (b, c) Mean shapes of asymmetric tricyclic chains at $\varepsilon$ = 10 for R160-R120-R80 and R200-R160-R120, respectively. (d) Plots of the number of closed rings against strain for each ring. When a ring satisfied the conditions $l_1 >$$l_{\rm min}$, $l_2 <$3, $l_3 <$3, and $l_4 <$3, it was deemed as being in the closed-ring state. The value $l_{\rm min}$ was determined from the steady-state behaviors of $l_1$, where $l_{\rm min}$ = 60 for $N_{\rm R}$ = 160, 40 for $N_{\rm R}$ = 120, and 20 for $N_{\rm R}$ = 80.}
    \label{fig12_assymmetric}
\end{figure}

In the present study, we investigated symmetric-multicyclic/linear blends under biaxial elongational flow and observed the stress overshoot induced by the topological transition in multicyclic chains with structural symmetry. 
To clarify the stress overshoot observed in asymmetric multicyclic chain, two asymmetric tricyclic chains, each composed of three different rings are discussed here, namely, R160-R120-R80 ($N_{\rm R}$ = 160, 120, and 80) and R200-R160-R120 ($N_{\rm R}$ = 200, 160, and 120). 
After preparing the asymmetric-tricyclic/linear blends with a 1:10 weight blend ratio, the blends were sufficiently equilibrated for 1.0 $\times$ 10$^9$ steps. 
Subsequently, a biaxial elongational flow with $\dot{\varepsilon}$ = 0.001$/\tau$ was applied. 
The obtained stress--strain curves are presented in Figure \ref{fig12_assymmetric}a and reveal that the asymmetric-tricyclic/linear blends do not exhibit stress overshoot. 
The mean shapes of R160-R120-R80 and R200-R160-R120 at $\varepsilon$ = 10 are presented in Figures \ref{fig12_assymmetric}b and \ref{fig12_assymmetric}c, respectively. 
In both cases, the largest rings are in open-ring states, while the middle-sized and smallest rings are in closed-ring states. 
The steady-state behaviors are similar to those observed for the symmetric-tricyclic chains. Because the open rings (the largest rings) mainly contributed to the steady-state stress, the steady-state value of R160-R120-R80/L40 corresponds to that of 3c-R160/L40. 
On the other hand, the steady-state value of R200-R160-R120/L40 is supposed to correspond to that of 3c-R200/L40, although this confirmation is beyond the scope of this study. 
For the asymmetric-tricyclic chains, the smaller rings preferentially adopt the closed-ring state under biaxial elongational flow (Figure \ref{fig12_assymmetric}d) because the number of linear chain penetrations is smaller. 
In contrast, in the symmetric-multicyclic chains, a ring takes the closed-ring state only when there is a difference in the number of linear chain penetrations between the interconnected rings. 
Otherwise, the rings in the symmetric-multicyclic chain elongate until a difference in the number of penetrations occurs. 
Specifically, the stress increases until a topological transition occurs. 
Structural symmetry in multicyclic chains is an important factor for inducing stress-overshoot behavior. 
Indeed, we supposed that a tricyclic chain composed of two symmetric large rings and a small ring, may exhibit stress-overshoot behavior. 
This hypothesis is verified in the next section.

\clearpage

\subsection{Tricyclic chains with two symmetric rings}\label{Sec3.5:TwoSymmetricRing}
\begin{figure}[htbp]
    \centering
    \includegraphics[width=7in]{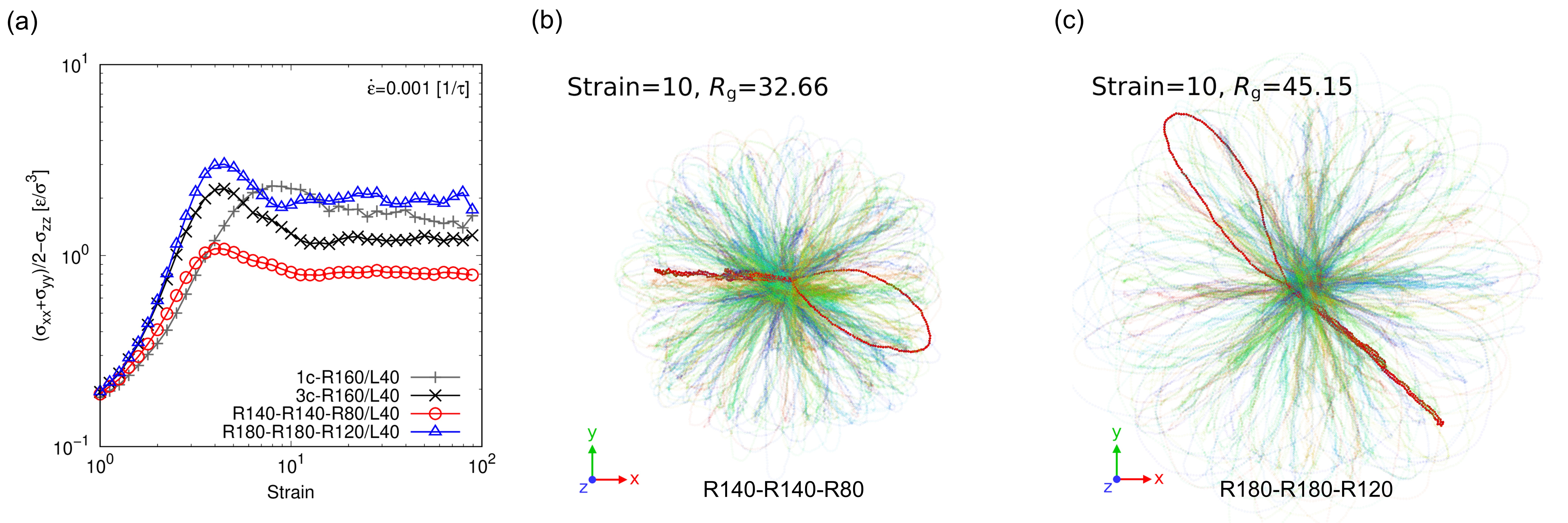}
    \caption{(a) Comparison of the stress–strain curves of R140-R140-R80/L40 and R180-R180-R120/L40 under biaxial elongational flow with a strain rate of 0.001$/\tau$ with those of 1c-R160/L40 and 3c-R160/L40. (b, c) Mean shapes of asymmetric tricyclic chains at $\varepsilon$ = 10 for R140-R140-R80 and R180-R180-R120, respectively.}
    \label{fig13_2large1small}
\end{figure}

The symmetric-tricyclic/linear blends exhibited stress-overshoot behaviors, whereas the asymmetric-tricyclic/linear blends did not. 
If we consider a tricyclic chain with two identical rings and one small ring, stress-overshoot phenomena are expected. 
We investigated two tricyclic chains with two symmetric rings, namely, R140-R140-R80 ($N_{\rm R}$ = 140, 140, and 80) and R180-R180-R120 ($N_{\rm R}$ = 180, 180, and 120). 
The obtained stress--strain curves of the two tricyclic chains under biaxial elongational flow with $\dot{\varepsilon}$ = 0.001$/\tau$ are presented in Figure \ref{fig13_2large1small}a. 
The tricyclic chains comprising two symmetric rings presented stress-overshoot phenomena, as expected. The mean shapes of R140-R140-R80 and R180-R180-R120 at $\varepsilon$ = 10 are shown in Figures \ref{fig13_2large1small}b and \ref{fig13_2large1small}c, respectively. 
The mean shapes represent genus 1, and the symmetry of the two symmetric rings is broken. 
The steady-state values of R140-R140-R80/L40 and R180-R180-R120/L40 are lower and higher, respectively, than those of 3c-R160/L40, and depend on the $N_{\rm R}$ of an open ring. 
Moreover, the value of the first normal stress difference at the peak was larger for a larger $N_{\rm R}$, while for $N_{\rm R} <$160, the strain at the peak was insensitive to $N_{\rm R}$. 
Further investigation is required for large rings ($N_{\rm R} >$160). 
These results show that multicyclic chains with symmetric rings exhibit stress-overshoot phenomena induced by the topological transition in the symmetric multicyclic chains. 
Thus, the symmetry of rings in a multicyclic chain is essential to produce stress overshoot under biaxial elongational flow.

\clearpage

\section{Summary and Future Outlook}
We investigated multicyclic/linear blends under biaxial elongational flow using coarse-grained molecular dynamics simulations. Our results revealed stress-overshoot phenomena in the multicyclic/linear blends, which were steeper than that previously reported for monocyclic/linear blends~\cite{MurashimaEtal2021}. 
Molecular morphology analyses revealed a {\it topological transition} mechanism in the symmetric-multicyclic chain. 
This transition is a morphological change from the open- to closed-ring state. 
We observed that in the closed-ring state, one ring was tightly closed to form a double-folded string. 
This double-folded string state, which was in a relaxed state under biaxial elongational flow, was not observed in the monocyclic chains. 
Because the closed-ring state inhibits chain penetration, the topological transition decreases the first normal stress difference under biaxial elongational flow. 
We therefore concluded that topological transition induces stress-overshoot phenomena. Comparison of the bicyclic/linear and tricyclic/linear blends revealed that the steady-state value of the first normal stress difference in the tricyclic/linear blends was smaller than that in the bicyclic/linear blends. 
This difference was also explained in terms of the topological differences between the bicyclic and tricyclic chains: Because the state of genus 1 (single open-ring state) is the most stable and populous state, the reduction in the first normal stress difference increases with the increasing number of rings in the multicyclic chain. 
In the case of tricyclic/linear blends, the state of genus 2 remains within a certain proportion, suggesting that the closure of one ring stabilizes this state. 
A multicyclic chain with more than three rings may represent a stable of genus 3 under biaxial elongational flow. Whether topological transitions occur under different flow conditions (e.g., shear flow and uniaxial elongational flow) remains an open query. Further investigations are currently underway to address these problems.

Stress overshoots were observed in the symmetric-multicyclic/linear blends. 
We also investigated asymmetric tricyclic chains and tricyclic chains composed of two symmetric rings and one small ring. 
The asymmetric tricyclic chains also showed morphological changes between the open- and closed-ring states, although the stress-overshoot behavior was inhibited in the asymmetric-multicyclic/linear blends. 
This is because smaller rings preferentially adopt closed-ring states. The blends of tricyclic chains with two symmetric rings and linear chains, however, displayed the stress-overshoot behavior. 
When a multicyclic chain has symmetric rings, there is no fixed order for the closed-ring states to occur, and thus,
structural symmetry in a multicyclic chain causes the dilemma of deciding which ring to close.
The strain on the rings increases until one ring is closed. 
This closed ring then relaxes, so that the ring stress is drastically decreased. 
Therefore, topological transition in symmetric-multicyclic chains induces the stress-overshoot behavior.

In the present study, we solely focused on the single strain rate value ($\dot{\varepsilon}=0.001 [1/\tau]$). The strain rate dependence is beyond the scope of this study and will be investigated in the future. Furthermore, we expect that the multicyclic/linear blends will present interesting phenomena under the different flow patterns (uniaxial, planar, and shear flows). Because the topological transition was caused by pulling out the linear chains penetrating the rings, we can regard this transition as a variant of the threading--unthreading transition~\cite{BorgerEtal2020}. The closed-ring state is apparently unthreaded by linear chains, and therefore, we did not investigate the number of penetrations~\cite{HagitaMurashima2021a,HagitaMurashima2021b,MurashimaEtal2021} in the present study. We expect that the multicyclic/linear blends also show such a threading–unthreading transition under uniaxial elongational flow. The topological transition, however, will be difficult to observe under uniaxial elongational flow, because both the threaded and unthreaded states will be tightly closed under such conditions. Recently, O’Connor et al. discussed the blend ratio dependence of monocyclic/linear blends under uniaxial elongational flow and revealed that the overshoot was remarkable at a certain ring volume fraction where the topological constraints of the composite entanglement network are maximized~\cite{OConnorEtal2022}. Indeed, the blend ratio dependence of the multicyclic/linear blends is also important; however, this will also be investigated in a future study. Notably, the open-closed state in a bicyclic/linear blend and open-closed-closed state in a tricyclic/linear blend under biaxial elongational flow correspond to the tadpole-shaped polymers~\cite{RosaEtal2020}. Comparison with the tadpole-shaped polymers is also a subject for future study.

In the present study, we considered a multicyclic chain in which multiple rings are connected by a bond for a bicyclic chain and three bonds for a tricyclic chain. 
As similar multicyclic chains, catenanes are also highly interesting~\cite{RauscherEtal2018,SawadaEtal2019a,SawadaEtal2019b}. 
In catenanes, interpenetration amongst multiple rings disrupt ring closure under biaxial elongational flow. 
Therefore, in the case of catenanes, the closed-ring state observed in the present work would be suppressed. 
We supposed that the catenane system does not present a topological transition. 
Recently, we discovered that a fluid-solid phase transition occurs as the catenane ring size is reduced~\cite{HagitaEtal2022e}. 
It has also been found that reducing the ring size increases the porosity. 
Thus, it is more difficult for a catenane to adopt the closed-ring state if the ring size is smaller. 
It is therefore expected that bond-connected multicyclic chains and catenanes display different characteristics under biaxial elongational flow. 
A comparison of the two materials is a topic for future research.

The present work suggests the possibility of tuning the physical properties by controlling the topology. 
The phenomenon observed in this work comprised a transition in which an open ring closes; however, the opposite process is also possible. 
For example, a ring closed by assuming an associative interaction~\cite{WuChen2022} can be opened by an external stimulus. 
Thus, a strategy to freely control the open and closed states of a ring would be useful for energy conservation, and the viscosity would be controlled by the topological transition. 
Moreover, we expect that this strategy will be used as a new method to recycle materials by tuning their topology. 
This work is just the beginning of the study of topological transitions. 
Future research will broaden our knowledge of topological transitions and bridge topological transitions to innovative applications.

Finally, we address the experimental verification possibilities in the present work. 
For the same ring size, the bicyclic/linear blends and the tricyclic/linear blends exhibited steeper strain hardening than did the monocyclic/linear blends. 
This behavior can be immediately useful as an indicator of multicyclization methodology. 
Figure \ref{fig7_transition} shows that the open-closed states account for approximately 10\% at $\varepsilon$ = 2. 
This suggests that the open-to-closed transition (or its transition process) can be observed by neutron scattering~\cite{Shibayama1998}. 
Under biaxial elongational flow, multicyclic chains orient randomly in the $xy$ plane (elongational plane). 
The closed-ring state has a rod-like shape, while the open-ring state displays a ring shape; thus, the form factors between the two states are different. 
Neutron scattering can therefore be used to detect and distinguish between the two states. 
The overshoot peak appeared at $\varepsilon\sim$4 and $\sim$8 in the multicyclic/linear and monocyclic/linear blends, respectively. 
Unfortunately, these overshoot behaviors under biaxial elongational flow are not observable with the current experimental technique, which is limited to $\varepsilon$ = 2.5~\cite{VenerusEtal2010,VenerusEtal2019}. 
However, we predict that future technological advances will eventually allow us to observe the overshoot behaviors under biaxial elongational flow, and thus, the overshoot of the multicyclic/linear blends can be further studied. 
The present work suggests that overshoot behavior appears at lower strain values when designing a method to connect the rings. 
Topological transitions in multicyclic chains with structural symmetry will play a key role in observing overshoot behavior under biaxial elongational flow in experiments.

\clearpage

\begin{acknowledgement}
TM thanks Prof. T. Taniguchi, Prof. M. Sugimoto, Prof. J.-I. Takimoto, Prof. S. K. Sukumaran, Prof. T. Uneyama, Prof. T. Honda, and Dr. Y. Tomiyoshi for their fruitful discussions, comments, and encouragement. The authors thank Prof. H. Jinnai, Prof. T. Satoh, and Prof. T. Deguchi for their support and encouragement. TM and TK thank the collaboration program of the Advanced Imaging and Modeling Center for Soft-materials (AIMcS) of Tohoku University. For the computations in this work, the authors were partially supported by the Supercomputer Center, Institute for Solid State Physics, University of Tokyo; MASAMUNE-IMR at the Center for Computational Materials Science, Institute for Materials Research, Tohoku University; Grand Chariot and Polaire at Hokkaido University Information Initiative Center; Flow at Nagoya University Information Technology Center; SQUID at Cybermedia Center, Osaka University; Fugaku at RIKEN Center for Computational Science; the Joint Usage/ Research Center for Interdisciplinary Large-scale Information Infrastructures (JHPCN); and the High-Performance Computing Infrastructure (HPCI) in Japan: hp200048, hp200168, hp210102, hp210132, hp220019, hp220104, hp220113, hp220114, jh210035, and jh220038. This work was partly financially supported by JSPS KAKENHI, Japan, Grant numbers JP18H04494, JP19H00905, JP20K03875, JP20H04649, and JP21H00111 and JST CREST, Japan, Grant numbers JPMJCR1993 and JPMJCR19T4. The authors thank Editage (\url{www.editage.com}) for the English language editing.
\end{acknowledgement}

\clearpage

\bibliography{MultiRingOvershoot}

\end{document}